\documentclass[twocolumn]{aa}  

\usepackage{multicol}
\usepackage{url}
\usepackage{longtable}
\usepackage{graphicx}
\usepackage{adjustbox}
\usepackage{hyperref}
\usepackage{xcolor}

\hypersetup{
 colorlinks = true,
  linkcolor = blue,
  anchorcolor = blue,
  citecolor = blue,
  filecolor = blue,
  urlcolor = blue%Colour of citations
}
\usepackage{txfonts}
\usepackage{amssymb}
\usepackage{esvect}
\usepackage{natbib,twoopt}
\usepackage{float}
\usepackage{color}

\usepackage{xcolor}
\begin{document}

   \title{The Fornax Deep Survey with VST
   }

   \subtitle{X. The assembly history of the bright galaxies and intra-group light in the Fornax~A subgroup}

   \author{ M. A. Raj\inst{\ref{inst1}, \ref{inst2}, \ref{inst3}}\and E. Iodice \inst{\ref{inst1}, \ref{inst3}} \and N.~R.~Napolitano\inst{\ref{inst4}} \and M. Hilker \inst{\ref{inst3}}  \and M.~Spavone \inst{\ref{inst1}} \and R. F. Peletier \inst{\ref{inst5}} \and H-S.~Su \inst{\ref{inst6}} \and   J.~Falc\'on-Barroso \inst{\ref{inst7}, \ref{inst8}} \and  G.~van de Ven  \inst{\ref{inst12}} \and M.~Cantiello \inst{\ref{inst9}} \and D.~Kleiner \inst{\ref{inst10}}\and A.~Venhola \inst{\ref{inst6}} \and S.~Mieske \inst{\ref{inst11}} \and M.~Paolillo \inst{\ref{inst2}, \ref{inst13}, \ref{inst1}}  \and M.~Capaccioli \inst{\ref{inst2}} \and P.~Schipani \inst{\ref{inst1}}}

\institute{INAF-Astronomical observatory of Capodimonte, via Moiariello 16, Naples, I-80131, Italy \\email: \texttt{mariaangela.raj@oacn.inaf.it}  \label{inst1}
   \and
   University of Naples ``Federico II'', C.U, Monte Santangelo, Via Cinthia, 80126, Naples, Italy  \label{inst2}
             \and
             European Southern Observatory, Karl-Schwarzschild-Strasse 2, D-85748 Garching bei Munchen, Germany \label{inst3}
             \and 
School of Physics and Astronomy,  Sun Yat-sen University Zhuhai Campus, 2 Daxue Road,  Tangjia,  Zhuhai,  Guangdong 519082,  P.R. China  \label{inst4}
\and
Kapteyn Astronomical Institute, University of Groningen, PO Box 72, 9700 AV Groningen, The Netherlands \label{inst5}
  \and 
Division of Astronomy, Department of Physics, University of Oulu, Oulu, Finland \label{inst6}
   \and
   Instituto de Astrof\`isica de Canarias, C Via Lactea s/n, 38200 La Laguna, Canary Islands, Spain \label{inst7}
   \and 
   Departamento de Astrof\`isica, Universidad de La Laguna, E-38200 La Laguna, Tenerife, Spain \label{inst8}
   \and 
   Department of Astrophysics, University of Vienna, T\"urkenschanzstrasse, 17, 1180 Vienna, Austria \label{inst12}
   \and 
   INAF-Astronomical Abruzzo Observatory, Via Maggini, 64100, Teramo, Italy \label{inst9}
   \and INAF - Osservatorio Astronomico di Cagliari, Via della Scienza 5, I-09047 Selargius (CA), Italy \label{inst10}
    \and
   European Southern Observatory, Alonso de Cordova 3107, Vitacura, Santiago, Chile \label{inst11}
   \and
   INFN - Sezione di Napoli, Napoli, 80126, Italy \label{inst13}
   }
   \date{Received 27 March 2020 ; Accepted 18 June 2020}

% \abstract{}{}{}{}{} 
% 5 {} token are mandatory
 
  \abstract
   {We present the study of the south-west group in the Fornax cluster centred on the brightest group galaxy (BGG) Fornax~A, which was observed as part of the Fornax Deep Survey (FDS). 
   This includes the analysis of the bright group members ($m_B < 16$ mag) and the intra-group light (IGL). }
   {The main objective of this work is to investigate the assembly history of the Fornax~A group and to compare its physical quantities as a function of the environment to that of the Fornax cluster core.}
   {For all galaxies, we extracted the azimuthally averaged surface brightness profiles in three optical bands ($g,r,i$) by modelling the galaxy's isophotes. We derived their colour ($g-i$) profiles, total magnitude, effective radius in all respective bands, stellar mass, and the break radius in the $r$-band. The long integration time and large covered area of the FDS allowed us to also estimate the amount of IGL.}
   {The majority of galaxies in the Fornax~A group are late-type galaxies (LTGs), spanning a 
   range of stellar mass of ${8 < \rm{log} (\mathit{M_* M_{\odot}}) < 10.5}$. Six out of nine LTGs show a Type~III (up-bending) break in their light profiles, which is either suggestive of strangulation halting star formation in their outskirts or their H\textsc{i}-richness causing enhanced star formation in their outer-discs. Overall, we do not find any correlations between their physical properties and their group-centric distance.  The estimated luminosity of the IGL is  $ 6 \pm 2 \times 10^{10} L_{\odot} $ in the $g$-band, which corresponds to about 16\% of the total light in the group. }
   {The Fornax~A group appears to be in an early-stage of assembly with respect to the cluster core. The environment of the Fornax~A group is not as dense as that of the cluster core, with all galaxies except the BGG showing similar morphology, comparable colours and stellar masses, and Type~III disc-breaks, without any clear trend in these properties with group-centric distances. The low amount of IGL is also consistent with this picture, since there were no significant gravitational interactions between galaxies that modified the galaxies' structure and contributed to the build-up of the IGL. The main contribution to the IGL is from the minor merging in the outskirts of the BGG NGC~1316 and, probably, the disrupted dwarf galaxies close to the group centre.  
   
   }

\titlerunning{Fornax~A subgroup}
\authorrunning{M.A.Raj et al. }
\maketitle
%
%-------------------------------------------------------------------

\section{Introduction}

Galaxies tend to gather in gravitationally bound systems during their evolution, which are seen as dense knots in the cosmic web filaments. In the Local Universe, it has been found that about half of the population of galaxies are found in groups and clusters \citep{Kara2005}. According to the hierarchical structure formation, these galaxy clusters are formed from the merging of smaller structures such as galaxy groups \citep[e.g.][]{blumen84, slarsen06, rudick09}. The terminology "galaxy group" and "galaxy cluster" depends on various factors, for example, the galaxy number density, the virial mass, X-ray luminosity, and velocity dispersion. Groups of galaxies or poor clusters have masses in the range from $10^{13}$ to $10^{14} M_{\odot}$ \citep{bower04,gary07}, and rich clusters are more massive $>10^{14} M_{\odot}$ \citep[e.g.][]{bahcall96}.  In addition to this, the intra-cluster medium (ICM) of rich clusters is hotter (X-ray temperature= 2--14 keV) than that of poor clusters and groups, and the radial velocity dispersion of galaxies in rich clusters is higher than that of groups \citep[e.g.][]{bahcall88, bahcall96}. 

The definition of these gravitationally bound systems range from binary galaxies to superclusters, all of which are important in the studies of environmental mechanisms. Several surveys have shown variations in observed properties of galaxies in different environments, that is, from the field to the core of clusters in the Local Universe \citep[e.g.][]{lewis02,kauff04, Peng10,grootes17, barsanti18} and also at higher redshifts, up to $z \sim 1$ \citep[e.g.][]{bianca08, lubin09, cucciati10, cooperm12, adam12, cucciati17, alfonso18, gugl19}. Such transitions in galaxy properties typically occur around cluster or group scales of $\sim 2\text{--}3$ virial radii \citep{gomez03}. One such observed property is the decrease in the star formation rate in galaxies of a given morphological type in groups and clusters  that is visible out to 2$r_{200}$ \citep{lewis02}. The mechanisms causing such changes are proposed to depend on the local density of the environment and, hence, they are different in groups and clusters.

The physical processes which are prominent in dense environments of clusters are ram-pressure stripping \citep{Gunn1972} and harassment \citep{Moore96}. Ram-pressure stripping is not prevalent in poor clusters and galaxy groups as it depends on the velocity of the galaxy and the hot ICM. Mergers and strangulation are more common in groups of galaxies \citep[e.g.][]{Barnes85, Zablu98}. Major mergers can significantly alter the morphology of spiral galaxies by destroying their discs and therefore producing a major remnant that resembles an elliptical galaxy \citep{dekel05}. Sometimes, the disc can survive gas-rich mergers albeit resulting in the build-up of a galactic bulge \citep[e.g.][]{stewart09,hopkins10}. On the other hand, disc strangulation is a result of the tidal forces in the group and cluster that suppresses the infall of gas onto the discs of galaxies which in turn reduces star formation \citep{larson1980}. The aforementioned processes that occur in groups of galaxies are most effective when the velocity dispersion of the group is similar to the internal velocity (rotation or dispersion) of its galaxy members \citep{Barnes85,Zablu98}. 

During the infall of galaxy groups which then form a cluster, the stripped material of galaxies settles to constitute the intra-group light (IGL) or the intra-cluster light  \citep[ICL;][]{nicola03, Yutaka2004, Contini14}. Therefore, the ICL and IGL are fossil records of past interactions and mergers as their growth and evolution over time provide a link to the assembly state of the cluster and group \citep[][and references therein]{mihos15}.

Overall, high density environments provide a plethora of information for galaxy formation and evolution. There have been several studies on the effect of the environment on galaxy structures and properties in the Local Universe \citep[e.g.][and references therein]{Dressler1980, Pasto94, demello02, lewis02, gomez03, bower04, Boselli2006b, gary07, huertas13, tully14, deel17, finn18} and at higher redshifts, up to $z \sim 1$ \citep[e.g.][and references therein]{cooper06,wel07,tasca09,nantais13,kelkar15,allen16, kry17, alf19}. With the advent of deep photometry \citep[e.g.][]{zibetti05}, exploring faint regions of galaxy outskirts, ICL and IGL have gained importance over the last years \citep[e.g.][and references therein]{rudick09, mihos17, Iodice2016, Spavone2017, Iodice2017b, Iodice2017, spavone18}. 

As such, the Fornax cluster has been an environment of interest. It is the second most massive galaxy concentration within 20 Mpc, after the Virgo cluster, with a virial mass of ${M_{\rm vir}= 7 \times 10^{13}}$~$M_\odot$ \citep{Drinkwater2001}. 
The Fornax cluster has been extensively studied by several surveys in multiple wavelengths 
\citep[e.g.][]{Martin2005,Scharf2005,Jordan2007,Davies2013,Munoz2015,Su2017,ngfs18,Zabel2018,serra19}.

The Fornax Deep Survey (FDS) with the VLT Survey Telescope \citep[VST,][]{Venhola2017, Iodice2019} 
is one of the deepest and widest data sets mapping the Fornax cluster out to the virial radius ($\sim0.7$~Mpc). 
With FDS we (i) mapped the surface brightness around the brightest cluster members NGC~1399 and NGC~1316 out to an unprecedented 
distance of about ${\sim 200}$~kpc ($R\sim6R_e$) and down to ${\mu_g \simeq 29\text{--}31}$~mag~arcsec$^{-2}$  \citep{Iodice2016,Iodice2017}; 
(ii) traced the spatial distribution of candidate globular clusters (GCs) inside $\sim0.5$~deg$^2$ of the cluster core \citep{Dabrusco2016,Cantiello2018}; (iii) studied the galaxy outskirts, detected the ICL and faint (${\mu_g \simeq28\text{--}30}$~mag~arcsec$^{-2}$) features in the intra-cluster region in the core of the cluster \citep{Iodice2016, Iodice2017b, Iodice2019, raj19} and in the outskirts of NGC~1316 \citep{Iodice2017}; (iv) provided the largest size and magnitude limited catalogue of dwarf galaxies in the cluster \citep{Venhola2017,Venhola2018,venhola2019}.
The analysis of the deep images from FDS suggests that the Fornax cluster is not completely relaxed inside the virial radius. The bulk of the gravitational interactions between galaxies takes place in the west-northwest core region, where most of the bright early-type galaxies (ETGs) are located and where the intra-cluster baryons (i.e. diffuse light, GCs, and PNe) are found  
\citep{Pota2018,Spiniello2018,Iodice2019}.

This paper focuses on the south-west (SW) Fornax~A group using the FDS data to study the bright group members ($m_B < 16$ mag) and their physical properties as a function of the environment and, thus, compare them with those observed for late-type galaxies (LTGs) in the Fornax cluster. \citet{Maddox2019} presented a spectroscopic census of all the galaxies in the Fornax environment, covered by FDS which also includes background objects. We use this data to study and compare the kinematics of the assembly history of the bright galaxies in the Fornax~A subgroup to those of the core.

The outline of this paper is as follows. In Sect.~\ref{fds3}, we give a brief summary of FDS. In Sect.~\ref{ana3}, we describe our methods of analysis along with the estimated parameters. In Sect.~\ref{results3}, we explain the results based on the surface photometry analysis. In Sect.~\ref{history}, we address the assembly history of the bright galaxies in Fornax~A. In Sect.~\ref{igl_disc}, we discuss the formation history of the IGL in the Fornax~A group. In Sect.~\ref{core_grp}, we compare the environments of the Fornax cluster core and Fornax subgroup.  In Sect.~\ref{conclu3}, we summarise our results. A brief description of each of the galaxies presented in this work is given in Appendix~\ref{ap_A} and their corresponding images ($g$-band), surface brightness profiles, ($g-i$) colour maps and colour profiles are shown in Appendix~\ref{sb_im}. Methodologies of some of this work are presented in Appendix~\ref{methods3}.

\section{The Fornax Deep Survey} \label{fds3}
The Fornax Deep Survey is a multi-band ($u,g,r,i$) survey, and a part of the Guaranteed Time Observation surveys FOCUS (P.I. R. Peletier) and VEGAS (P.I. E.Iodice, \citealt{Cap2015}). The observations were acquired with the ESO VLT Survey Telescope (VST), located at Cerro Paranal,  which is a 2.6-metre optical telescope \citep{Schipani2012} equipped with the wide field camera OmegaCam \citep{Kui2011}, that covers 1$\times$1 deg$^2$ field of view. FDS data sample 26 deg$^2$ of the Fornax cluster, which also includes the in-falling 
Fornax~A subgroup \citep{Iodice2017}. Detailed explanations of the observing strategy and reduction are given by \citet{Iodice2016, Iodice2017b} and \citet{Venhola2018}. The FWHM (arcsec) and depth (mag arsec$^{-2}$) of the FDS fields in each filter ($g, r, i$), used in this work, are given in Table~\ref{tab_psf} \citep[see also][]{Venhola2018}.

\begin{figure*}
   \centering
   \includegraphics[scale=0.24]{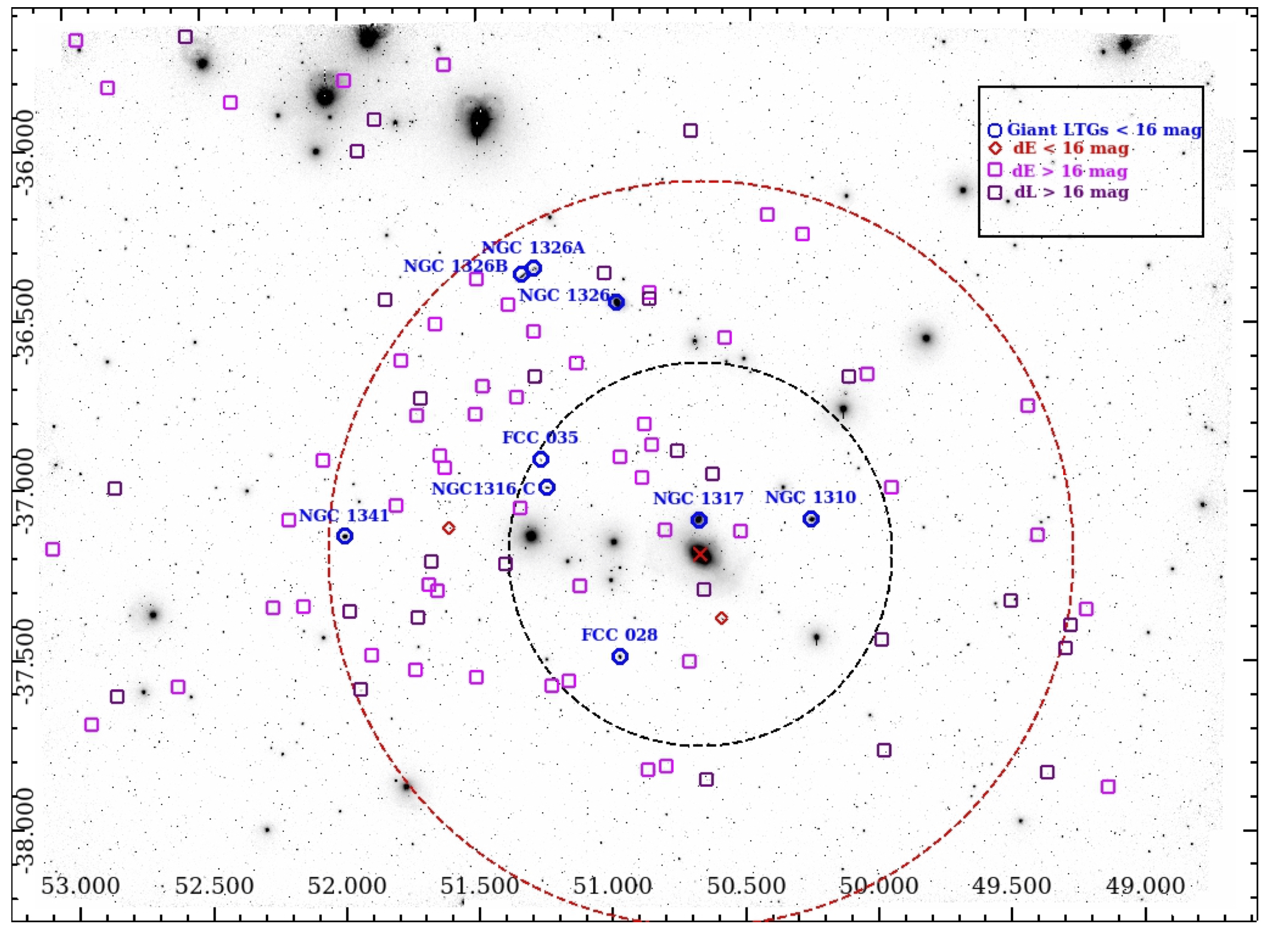}
      \caption{VST mosaic in $g$-band of the Fornax~A subgroup covering 3 $\times$ 2 deg$^2$. The red dashed circle is an approximate measure of the virial radius of the group, which is $\sim$1 deg$^2$ \citep{drinkwater2001b}. The black dashed circle ($r=0.58$ deg) is the region where the IGL is estimated. All galaxies part of the Fornax cluster and Fornax~A subgroup, taken from \citet{venhola2019} are indicated. The red cross indicates NGC~1316. Bright galaxies ($m_B<16$~mag) are represented with blue circles. Red diamond represents dE  ($m_B< 16$~mag), pink squares represent dE ($m_B>16$~mag), and purple squares represent dl ($m_B>16$~mag).}
           
         \label{group_g}
   \end{figure*}

Catalogues and data products provided by FDS were presented by \citet{Venhola2018} on dwarf galaxies 
in the Fornax cluster, \citet{Iodice2019} on photometry of ETGs, and \citet{raj19} on LTGs in the central 9 deg$^2$. The $g$-band mosaic covering 3.6$\times$2 deg$^2$ around NGC~1316 and the SW side of the Fornax cluster 
with named circles indicating the galaxies studied in this work, is shown in Fig.~\ref{group_g}. 
In this image, we also mark all galaxies taken from the catalogue by  \citet{venhola2019} that are part of the Fornax~A subgroup for the purpose of group-environmental analysis. The galaxies marked in Fig.~\ref{group_g} are grouped into giant LTGs ($m_B < 16$ mag), dwarf elliptical (dE), and dwarf late-type (dl) galaxies. The sample of spiral galaxies ($m_B < $16 mag) has been selected from \citet{Ferguson1989} and are listed in Table~\ref{tab2}. 
In this study, we adopt a distance of $D=20.8 \pm 0.5$ Mpc for NGC~1316 \citep{mik13}.  

\begin{table*}[h]
\caption{Image quality of FDS fields used in this work.}              % title of Table
\label{tab_psf}      % is used to refer this table in the text
\centering                                      % used for centering table
\begin{tabular}{cccccccc}          % centered columns (4 columns)

\hline\hline                        % inserts double horizontal lines
Field 
& \multicolumn{2}{c}{$g$ band } 
& \multicolumn{2}{c}{$r$ band}
& \multicolumn{2}{c}{$i$ band}
  % table heading
\\[+0.1cm]
\hline 
 &\multicolumn{1}{c}{$FWHM$}
 &\multicolumn{1}{c}{depth} 
 &\multicolumn{1}{c}{$FWHM$}
 &\multicolumn{1}{c}{depth} 
 &\multicolumn{1}{c}{$FWHM$}
 &\multicolumn{1}{c}{depth}\\

 & [arcsec]
 & [$\frac{mag}{arcsec^2}$]
 & [arcsec]
 & [$\frac{mag}{arcsec^2}$]
 & [arcsec]
 & [$\frac{mag}{arcsec^2}$]
\\[+0.1cm]
 (1)
  &\multicolumn{2}{c}{(2)}
  &\multicolumn{2}{c}{(3)}
  &\multicolumn{2}{c}{(4)}\\
\hline \hline
F22 & 1.04 $\pm$ 0.07 & 26.52 & 0.81$\pm$ 0.05 & 25.90 & 0.85 $\pm$ 0.07 & 25.16\\
F25 & 1.11 $\pm$ 0.10 & 26.63 & 0.77$\pm$ 0.06 & 25.84 & 0.85 $\pm$ 0.07 & 25.11\\
F26 & 0.93 $\pm$ 0.07 & 25.89 & 0.81$\pm$ 0.05 & 25.96 & 0.91 $\pm$ 0.07 & 25.06\\
F27 & 1.07 $\pm$ 0.10 & 26.39 & 0.78$\pm$ 0.06 & 25.63 & 0.89 $\pm$ 0.10 & 24.88\\
F28 & 1.08 $\pm$ 0.14 & 26.31& 0.79$\pm$ 0.06 & 25.57 & 0.92 $\pm$ 0.09 & 24.89\\

\hline
\end{tabular}
\tablefoot{Column 1 -- FDS fields ; Column 2 to 4 -- average seeing and rms of the FWHM, and surface brightness corresponding to 1$\sigma$ S/N per arcsec$^2$ for $g, r, i$ bands. The PSF FWHM and depth of each filter in all fields are adopted from \citet{Venhola2018}.} 
\end{table*}

\begin{table*}[h]
\caption{LTGs brighter than $m_B < 16$ mag in the Fornax~A subgroup.}              % title of Table
\label{tab2}      % is used to refer this table in the text
\centering                                      % used for centering table
\begin{tabular}{cccccccc}          % centered columns (4 columns)
\hline\hline                        % inserts double horizontal lines
FCC name & $\alpha$ & $\delta$  & Morph. type & Radial velocity
  & $m_B$ & FDS Field & Alternative names \\    % table heading
            &h:m:s&d:m:s& &km s$^{-1}$&mag& & \\
  (1)&(2)&(3)&(4)&(5)&(6)&(7)& (8)\\
 \hline\hline
FCC013 & 03 21 03.4	& $-$37 06 06 & SBcII     &  1805   & 12.55   &    F28     &  NGC~1310, ESO 357-G 019\\
FCC022 & 03 22 44.3& $-$37 06 13	&  Sa pec  &  1941  & 11.91       &    F26   & NGC~1317, ESO 357-G 023\\
FCC028 &03 23 54.2&$-$37 30 33&SmIII&1405&13.88&F26& ESO 301-IG 011\\
FCC029 & 03 23 56.4 & $-$36 27 53	&  SBa(r)	          &  1360   & 12.20 &    F25   &  NGC1326, ESO 357-G 026\\
FCC033 &03 24 58.4 & $-$37 00 34& SdIII pec & 1800&14.32& F26& NGC~1316C, ESO 357-G 027\\ 
FCC035 & 03 25 04.2 & $-$36 55 39	&  SmIV         &  1800&    15.3  &    F26            &  \\
FCC037 &03 25 08.5& $-$36 21 50 &SBcIII &1831 &13.77&F25&NGC~1326A, ESO 357-IG 028\\
FCC039 & 03 25 20.3 & $-$36 23 06&  SdIII	          &  999    &   13.59       &     F25           & NGC~1326B, ESO 357-G 029\\
FCC062 & 03 27 58.4 & $-$37 09 00&  SbcII	          &  1876  &  13.21              & F22 & NGC~1341, ESO 358-G 008\\
\hline
\end{tabular}
\tablefoot{Column 1 -- Fornax~A subgroup members from \citet{Ferguson1989}; Columns 2 and 3 --  Right Ascension and Declination (J2000); Column 4 -- Morphological type from \citet{Ferguson1989}; Column 5 -- Heliocentric radial velocity obtained from NED; Column 6 -- Total magnitude in $B$ band as given in NED (The NASA/IPAC Extragalactic Database (NED) is operated by the Jet Propulsion Laboratory, California Institute of Technology, under contract with the National Aeronautics and Space Administration.); Column 7 -- Location in the FDS Field; Column 8 --  Alternative catalogue names.} 
 \end{table*}
 
\section{Analysis} \label{ana3}
In this section, we briefly describe the procedures to derive galaxy parameters. We adopt the methods presented by \citet{Iodice2019} and \citet{raj19}. 
Results for individual galaxies of the sample are given in Appendix~\ref{ap_A}. 
\subsection{Isophote fitting} \label{iso}
The azimuthally averaged surface brightness (SB) profiles of each source were extracted in three bands ($g, r, i$) using the IRAF task \textsc{ellipse}\footnote{IRAF is distributed by the National Optical Astronomy Observatories, which are operated by the Association of Universities for Research in Astronomy, Inc., under cooperative agreement with the National Science Foundation} \citep{ellipse87}. 
 As preliminary steps to the isophote fitting, we masked all bright sources (e.g. galaxies, stars) that are close to the galaxies in the sample. In some cases, for galaxies that are close in projection to each other (NGC~1326A and NGC~1326B; NGC~1316 and NGC~1317), we adopted an iterative process by modelling the neighbouring source using the \textsc{bmodel} task in IRAF. This model is created from the results of the isophotal analysis (\textsc{ellipse}) and then subtracted from the original image.
In the fitting routine, we adopted the k-sigma clipping algorithm for cleaning deviant sample points at each annulus\footnote{Overall, the manual masking, 3$\sigma$ clipping and model-subtraction improve the isophotal fitting, ensuring the rejection of unmasked sources and defects.}.

Isophotes are fitted in elliptical annuli, starting from the centre\footnote{In the dust-free galaxies, the centre corresponds to the intensity peak. If the galaxy light is strongly affected by the dust absorption, we adopted the geometric centre of the isophotes.} of the galaxy to the edge of the FDS field ($\sim 0.5$ deg). Since most of the galaxies have dust absorption and/or irregular shape close to their centre, we keep the geometrical centre fixed, while the position angle and ellipticity are free parameters. 
According to \citet{Iodice2016}, the intensity profiles are then used to derive (i) the limiting radius ($R_{lim}$) that corresponds to the outer-most annulus, where the light of the source blends into the average background level (it is the residual after subtracting the sky frame resulting in a value close to zero); (ii) derive the residual sky background of each source from their outer annuli in three respective bands.  The limiting radius and the residual background levels are locally estimated for each galaxy in the sample. The background levels are computed at $R\geq R_{lim}$ as the average counts out to the edge of the image.

\begin{table*}[h]
\caption{Derived parameters of LTGs in the Fornax~A subgroup.}              % title of Table
\label{tab:param}      % is used to refer this table in the text
\centering                                      % used for centering table
\begin{tabular}{lcccccccc}          % centered columns (4 columns)
\hline\hline                        % inserts double horizontal lines
Object & $m_g$ & $g-r$ & $g-i$ & $ Re_r$ & $D_{core}$ & $M_i$ & $M_*$ & $M/L$ \\   
        & mag  & mag   & mag & arcsec & deg & mag & $10^{10}$~$M_{\odot}$ &  \\
  (1)&(2)&(3)&(4)&(5)&(6)&(7)&(8)&(9)\\
 \hline\hline
FCC013 (NGC~1310)& 12.48 $\pm$0.04& 0.57$\pm$0.1 & 0.77$\pm$0.02 & 27.6$\pm$0.2& 0.25& $-$19.74 & 0.47 & 0.76\\
FCC022 (NGC~1317)& 11.15$\pm$0.01 & 0.77$\pm$0.02 & 0.99$\pm$ 0.02 & 35.36$\pm$0.03& 0.02& $-$20.98 &1.71 & 0.88\\
FCC028 &13.33$\pm$0.02 & 0.57$\pm$0.04 & 0.98$\pm$0.05 & 22.1$\pm$0.5& 0.28& $-$19.07 &0.29 & 0.87\\
FCC029 (NGC~1326)& 10.62$\pm$0.02& 0.62$\pm$0.04 & 1.01$\pm$0.04& 48.2$\pm$1& 0.68& $-$21.53 &2.94 & 0.91\\
FCC033 (NGC~1316C)&13.81$\pm$0.03 &0.69$\pm$0.1 &0.97$\pm$0.04& 22.6$\pm$0.5& 0.37& $-$18.30 &0.14 & 0.85\\
FCC035 &14.79$\pm$0.03& 0.17$\pm$0.1 & 0.18$\pm$0.1 & 17.1$\pm$0.4 & 0.44& $-$17.39&0.017 & 0.24\\
FCC037 (NGC~1326A) & 13.55$\pm$0.03 & 0.51$\pm$0.1 & 0.95$\pm$0.1 & 41.0$\pm$0.8& 0.87& $-$18.54 & 0.17 & 0.83\\
FCC039 (NGC~1326B)&12.87$\pm$0.1& 0.32$\pm$0.1 & 0.73$\pm$0.1 & 53.03$\pm$2& 0.86& $-$19.00 &0.18 & 0.58\\
FCC062 (NGC~1341)&12.37$\pm$0.02 & 0.54$\pm$0.04& 1.01$\pm$0.04& 20.6$\pm$0.4& 0.92& $-$19.72 &0.55 & 0.91\\
\hline
\end{tabular}
\tablefoot{Column 1 -- Fornax~A subgroup members from \citet{Ferguson1989}; Column 2 -- Total magnitude in $g$-band derived from the isophote fit. Values were corrected for the Galactic extinction, using the absorption coefficient by \citet{Schlegel98}; Columns 3 and 4 -- Average $g-r$ and $g-i$ colours; Column 5 -- Effective radius in $r$-band, in arcsec; Column 6 -- Projected distance from the galaxy centre in degree, that is, from NGC~1316 (FCC021); Column 7 -- Absolute magnitude in $i$-band, derived using the distance modulus from NED and \citet{Tully2009}; Columns 8 and 9 -- Stellar mass and mass-to-light ($M/L$) in $i$-band.}
\end{table*}

\subsection{Products: Total magnitude, effective radius, colours, and stellar mass }
For each galaxy of the sample, we derived the "total magnitude", "effective radius" $R_e$, average "$g-i$" and "$g-r$" colours following the method presented by \citet{Iodice2019} and \citet{raj19}. All parameters are given in Table~\ref{tab:param}. Firstly, the intensity profiles from the output of the isophote fit are used to derive azimuthally averaged SB profiles. The residual background level estimated at $R \geq R_{lim}$ of each band for all galaxies is subtracted. We also took into account the uncertainties on the photometric calibration and sky subtraction in estimating the errors on magnitudes \citep[see][]{Cap2015, Iodice2016, Iodice2019}. For all galaxies in the sample, we derived the $g-i$ colour profiles from the SB profiles, and colour maps from the images, which are shown in Appendix~\ref{sb_im}. We then derived total magnitudes and effective radii in the $g, r, i$ bands of all galaxies in our sample using the growth curve from the isophote fits.  We estimated the stellar mass $M_{*}$ using the empirical relation from \citet{Taylor2011} which assumes a Chabrier IMF, that is,  ${\rm{log}_{10} \frac{\mathit{M_*}}{\mathit{M_{\odot}}} = 1.15 + 0.70 (\mathit{g-i}) - 0.4\mathit{M_i}}$.  Here $M_i$ is the absolute magnitude in $i$-band, $g-i$ is the average colour derived in the previous step. We then estimated the stellar mass-to-light ratio $M_*/L_i$ with 1$\sigma$ accuracy of $~$0.1 dex \citep[see][]{Taylor2011}.

\subsection{Estimation of the IGL}\label{sec:IGL_method}
The IGL is estimated from the mosaic in the $g$ and $r$ bands as their surface brightness limit is fainter than that of the $i$-band (see Table~\ref{tab_psf}). The brightest stars and galaxies surrounding NGC~1316 are modelled using the IRAF tasks \textsc{ellipse} and \textsc{bmodel} as explained in Sect.~\ref{iso}, and then subtracted from the image. The resultant residual image is used to derive the radial profiles and thus 2D model of NGC~1316, which is also subtracted from the image. The final residual image is used to estimate the IGL using the IRAF task \textsc{phot}, which computes the integrated flux inside a specified aperture. The uncertainties on the estimate of the IGL take into account 1) the error on the background estimate ($\simeq 10\%$ in $g$, $\simeq 50\%$ in $r$); 2) the poissonian error on the flux ($\simeq 26\%$ in $g$ and $r$) and 3) the error on the photometric calibration ($\simeq 0.1\%$ in $g$ and $r$). 

According to \citet{Iodice2016}, the stellar envelope of NGC~1316 starts to dominate at $R >R_{tr}= 5.5^{\prime}$, where $R_{tr}$ is the transition radius\footnote{The transition radius $R_{tr}$ is defined as the distance from the centre of the galaxy to the region where an inflection (evident in their SB profiles) in the stellar halo occurs \citep{cooper10, cooper13}.}. Taking this into account, we masked this region of NGC~1316. The IGL is estimated in a region $\sim$33.5 arcmin$^2$ surrounding NGC~1316, which covers nearly 60\% of the group including the bright galaxies. In Appendix. \ref{igl_image}, we show the final $g$-band image used in estimating the IGL of the Fornax~A subgroup. In this image, we also mark the loops that are part of the stellar envelope (see also \citealt{Iodice2017}). Through our deep photometry  of this group, we also find a new patch of diffuse light in the SW part of NGC 1316, which also contributes to the IGL (see Appendix.~\ref{igl_image}). 

The total magnitude\footnote{Values were corrected for Galactic extinction using the absorption coefficient by \citet{Schlegel98}.} of the IGL in the Fornax~A group, within an area of $\sim$33.5 arcmin$^2$ around NGC~1316, is ${m_g=10.1\pm0.3}$~mag and ${m_r=9.2\pm0.3}$~mag in the $g$ and $r$ bands, respectively. Therefore, the total luminosity is ${6 \pm 2 \times 10^{10}L_{\odot}}$ in the $g$ band and ${7 \pm 2 \times 10^{10}L_{\odot}}$ in the $r$ band. As pointed out earlier, the IGL estimate includes the stellar envelope of NGC~1316. The fraction of the IGL luminosity to the (i) luminosity of the brightest group galaxy (BGG) NGC~1316 is $L_{IGL,g}/ L_{BGG,g} \sim 19\%$; (ii) total luminosity of all group members (including dwarf galaxies) inside the virial radius is $L_{IGL,g}/ L_{group,g}\sim 16 \pm 4 \%$. The total luminosity of NGC~1316 ($L_{BGG,g}=3.11 \times 10^{11} L_{\odot}$) is given by \citet{Iodice2017}. The total luminosity for dwarf galaxies ($9\times10^{8} L_{\odot}$ in $g$-band) is provided by \citet{venhola2019}. The total luminosity for all bright LTGs in the sample studied in this work are derived from magnitudes listed in Table~\ref{tab:param}.  

The contribution to the IGL (in $g$-band) of the luminous loops in the stellar envelope of NGC~1316 (loops L1, L2, L9, SW loop L5) is  ~22\%. According to \citet{Iodice2017}, the total luminosity from these loops is $ 1.3 \times 10^{10} L_{\odot}$ in the $g$-band. We stress that the IGL estimation is a lower limit as it is not possible to estimate the diffuse component which is in the line-of-sight of masked objects.

\begin{table*}[h]
\caption{Parameters of LTGs with a disc break.}              % title of Table
\label{tab4}      % is used to refer this table in the text
\centering                                      % used for centering table
\begin{adjustbox}{width=19 cm,center}
\begin{tabular}{lcccccccccc}          % centered columns (4 columns)

\hline\hline                        % inserts double horizontal lines
Object 
& \multicolumn{2}{c}{$B_r $}
&$\sigma_{nB_{r}}$   
&$\mu_{B_{r}}$     
& \multicolumn{2}{c}{$  h_{\rm{in}} $ } 
& \multicolumn{2}{c}{$ h_{\rm{out}}$ }
& Break    % table heading
& Hubble
\\[+0.1cm]
 &\multicolumn{1}{c}{[arcsec]}
 &\multicolumn{1}{c}{[kpc]} 
 &  [arcsec]                              
 & [$\frac{mag}{arcsec^2}$]
 &\multicolumn{1}{c}{[arcsec]}
 & \multicolumn{1}{c}{$g-i$ [mag]}
 & \multicolumn{1}{c}{[arcsec]}
 &\multicolumn{1}{c}{$g-i$ [mag]}
 & Type
 & Type ($T$)
\\[+0.1cm]
  (1)
  & \multicolumn{2}{c}{(2)}
  &\multicolumn{1}{c}{(3)}
  &\multicolumn{1}{c}{(4)}
  &\multicolumn{2}{c}{(5)} 
  &\multicolumn{2}{c}{(6)}
  &\multicolumn{1}{c}{(7)}
   &\multicolumn{1}{c}{(8)} \\

  \hline \hline
FCC013(NGC~1310) &  76.8 & 7.76 &3.7 & 25.01 &  54.2 &0.83$\pm$0.03 & 43.7  & 1.27 $\pm$ 0.1  &III & 5\\
FCC022(NGC~1317)  &   - &    &-   & - &  -      & - & -  &    &I  & 1\\
FCC028  &   38.8 &  3.91 &1.6 & 23.60  &   24.0  &  0.83 $\pm$ 0.1& 60.8   &   0.87 $\pm$ 0.1   &III & 9\\
FCC029(NGC~1326)  &   53.6  & 5.41 & 1.5  &  21.82    &24.76& 0.92 $\pm$ 0.07& 159.8 & 0.68 $\pm$ 0.03  & III  & 1\\
FCC033(NGC~1316C) &   - &    &-   & - &  -      & - & -  &    & I & 7\\
FCC035  &27.6 & 2.8& 0.3& 23.94&  25.64 & $-$0.57 $\pm$ 0.02 &34.2  &  $-$0.93 $\pm$ 0.1 &III &9\\
FCC037(NGC~1326A)  &   - &    &-   & - &  -      & - & -  &    &I&5 \\
FCC039(NGC~1326B)& 95.5 & 9.64& 8.2 & 25.01 & 94.9 & 0.52$\pm$ 0.02 & 37.0& 0.88 $\pm$ 0.2 &III&7\\
FCC062(NGC~1341)  &   52.3  & 5.28& 2.3 & 24.03 &  38.9 &0.78$\pm$ 0.02 & 68.2  & 0.86 $\pm$ 0.03 &III&4\\

\hline

\end{tabular}
 \end{adjustbox}
\tablefoot{Column 1 -- LTGs with a disc break; Column 2 -- Break radius in units of arcsec and kpc (1 arcsec $=$ 0.101 kpc);  Column 3 -- Standard deviation of the break radii from $(n+1)^3$ combinations; Column 4 -- Surface brightness at the break radius; Column 5 -- Inner scale-length in units of arcsec, and average $g-i$ colour for $h_{\rm{in}}$; Column 6 --  Outer scale-length in units of arcsec, and average $g-i$ colour for $h_{\rm{out}}$; Column 7 -- Profile classification; Column 8 -- Hubble Type $T$ taken from \citet{drinkwater01}.} 
 \end{table*}
 
\subsection{Products: Break radius, disc scale-length, and disc colour} \label{algor}
We used the algorithm developed by \citet{raj19} to derive the break radius $B_r$, inner and outer scale-lengths ($h_{\rm{in}}$ and $h_{\rm{out}}$)\footnote{The disc scale-length refers to the region of the disc before and after the break radius.} from the deconvolved SB profiles in $r$-band. The ($g-i$) colours of the scale-lengths ($h_{\rm{in}}$ and $h_{\rm{out}}$) are also derived from deconvolved SB profiles. The deconvolution method is briefly described in the following. 

\begin{figure*}
   \centering
   \includegraphics[width=17cm]{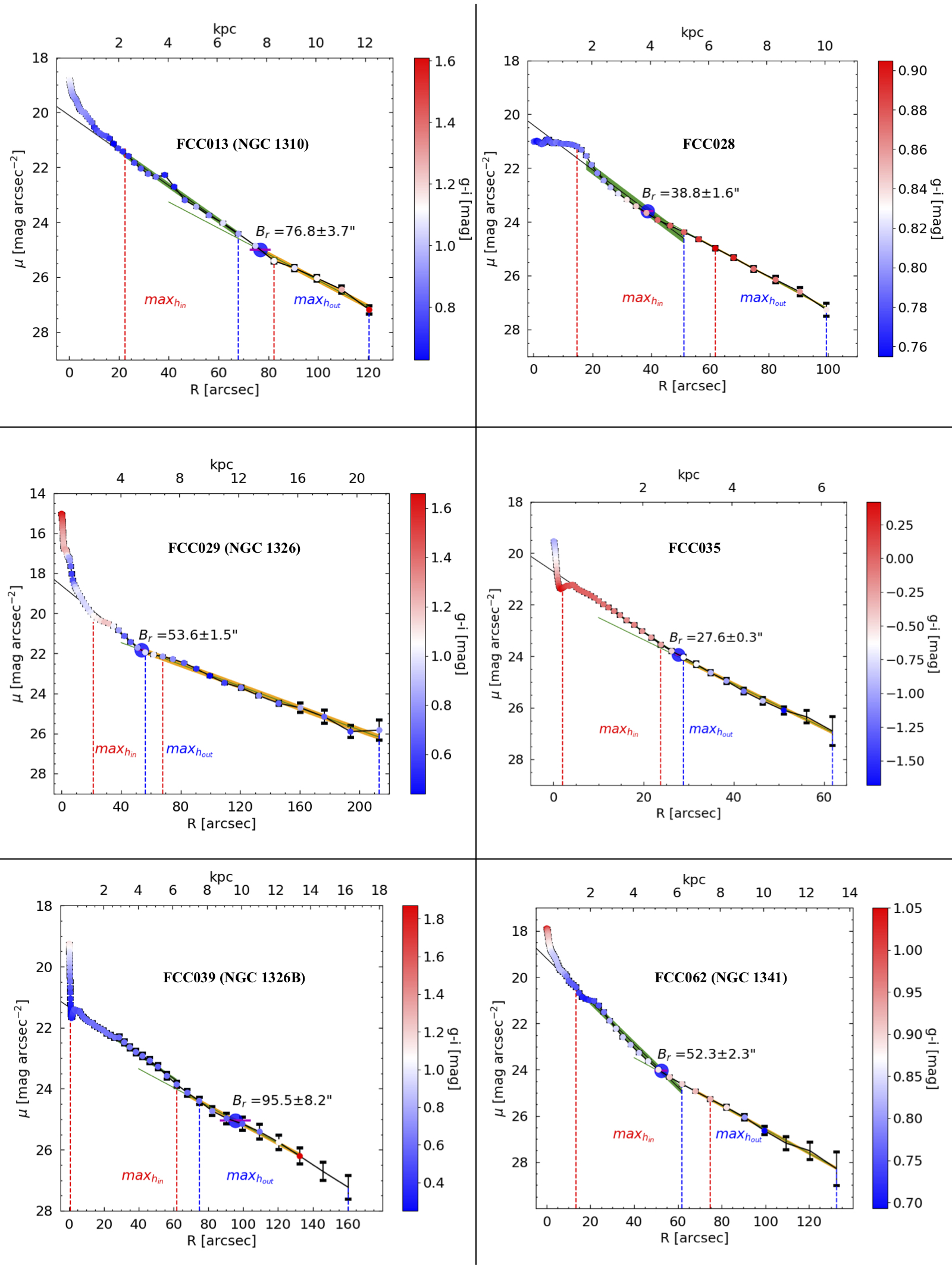}
      \caption{Surface brightness profiles of galaxies with Type~III disc break, with ($g-i$) colour as a third parameter. In each plot, the break radius ($B_r$) is marked at the intersecting point of the linear fits performed between $h_{\rm{in}}$ and $h_{\rm{out}}$, with $\sigma_{nB_r}$ as the median of $(n+1)^3$ combinations of best fits on the inner and outer scale-lengths. The vertical dashed lines ($\rm{max}_{\textit{h}_{\rm{in}}}$ and $\rm{max}_{\textit{h}_{\rm{out}}}$) indicate the regions for $\rm{max_{range_{in}}} = \rm{range_{in}} \pm \textit{n} $ and $\rm{max_{range_{out}}} =  \rm{range_{out}} \pm \textit{n}$ where the algorithm produces $(n+1)^2$ linear least square fits. The shaded regions on $h_{\rm{in}}$ and $h_{\rm{out}}$ indicate the median of the  rms of the residuals for $(n+1)^2$ linear least square fits. }
             \label{truncs}
   \end{figure*}

In the process of detecting disc-breaks, we made sure to account for the effect of the point spread function (PSF). This is to avoid false detection of disc-breaks as a result of the scattered light from a galaxy's bright core in the regions around them. We first measured the PSF out to the galaxy's disc and then the PSF is characterised from the VST images so that the broadening effect of the seeing on the galaxies is taken into account \citep{Cap2015}. \citet{raj19} show how the effect of the PSF can affect the SB profiles of galaxies of various sizes \citep[see also][]{Cap2015}. We then deconvolved galaxies with the PSF, using the Lucy-Richardson algorithm \citep{Lucy74, Richardson72}. We briefly explain the main steps of the algorithm used to derive the break radius. 
\begin{enumerate}
    \item The disc regions are determined from the 1D multi-component (bulge and disc) fits explained in Appendix~\ref{multi} (the multi-component best fits are shown in Fig.~\ref{fits} and their corresponding best fit parameters are listed in Table~\ref{best_fit}). The starting point is the intersection between the bulge and disc components. The regions are provided to the algorithm to define inner disc limit $\rm{range_{in}}$ and outer disc limit $\rm{range_{out}}$, within $R_{lim}$ (see Table~\ref{tab_lim}). 
    \item The algorithm produces $(n+1)^2$ best linear fits by varying $\rm{range_{in}}$ and $\rm{range_{out}}$, and $(n+1)^3$ estimates of the break radius, where $n=2$ in both cases.  
    \item The median of the $(n+1)^3$ intersecting points is chosen as the final break radius $B_r$, and the standard deviation of the $(n+1)^3$ estimates is given as sigma $\sigma_{nB_r}$ on the estimation of the break radius. 
    \item Through iterations\footnote{The number of iterations varies for each galaxy with a minimum of five until a minimal standard deviation of the $(n+1)^3$ break radii is obtained.} of the previous steps, a minimal standard deviation of the $(n+1)^3$ break radii is obtained which in turn provides the best selection of the disc scale-lengths $h_{\rm{in}}$ and $h_{\rm{out}}$. 
    \item The average ($g-i$) colours of the inner and outer discs are derived for the regions $h_{\rm{in}}$ and $h_{\rm{out}}$. 
\end{enumerate}
The break radius (in arcsec and kpc) of six galaxies along with $\sigma_{nB_r}$, surface brightness at the break $\mu_{B_r}$, and average $(g-i)$ colour for the components $h_{\rm{in}}$ and $h_{\rm{out}}$ are listed in Table~\ref{tab4}. Fitting results are shown in Fig.~\ref{truncs}.

\section{Results} \label{results3}
\subsection{Morphology and colour analysis of galaxies in the Fornax~A subgroup}
In this section, we analyse the morphology, structure, colours, and stellar mass. as a function of group-centric distance, that is from BGG NGC~1316. This choice is motivated by making a direct comparison with the segregation of the galaxies' properties found in the cluster core \citep{Iodice2019,raj19}, as discussed in Sect.\ref{core_grp}.

\begin{figure*}
%\hspace{-0.55cm}
% \vspace{-0.25cm}
   \includegraphics[scale=0.25]{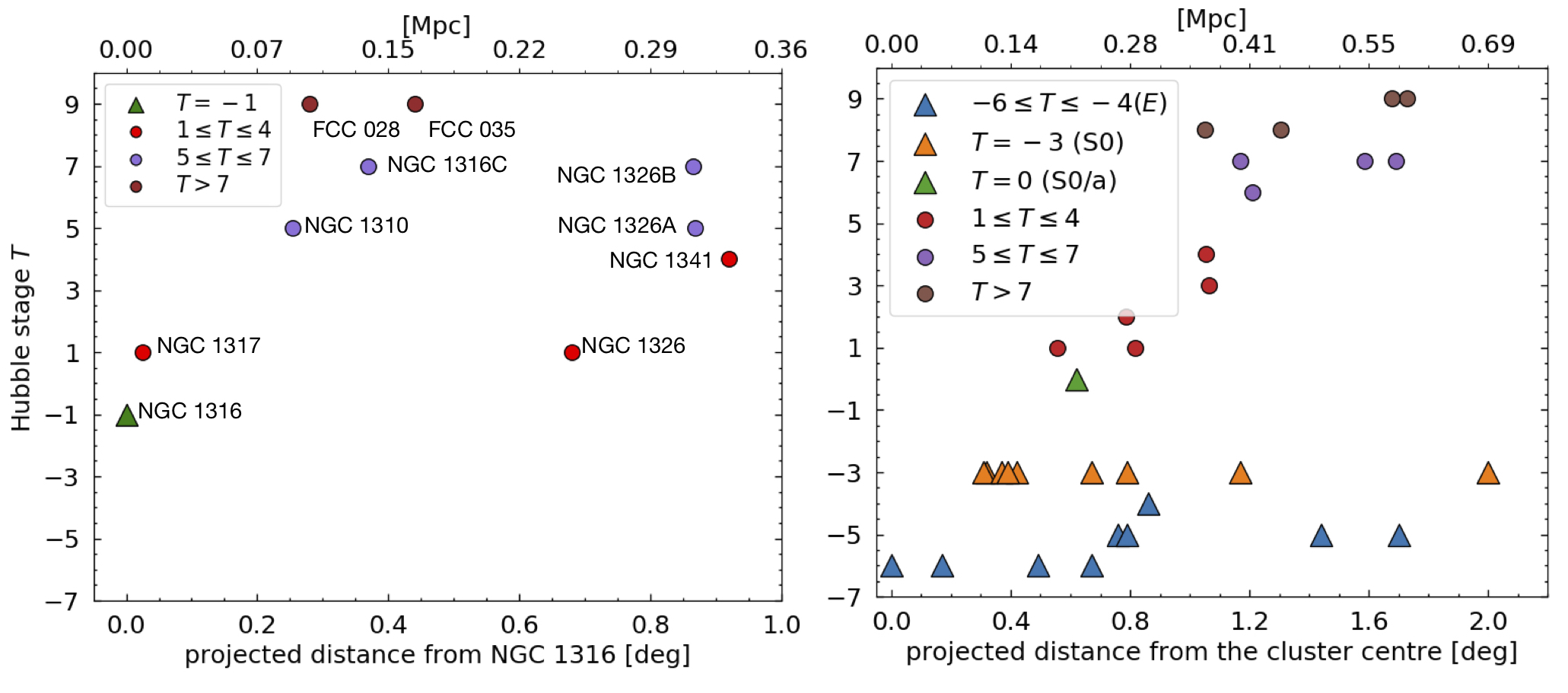}
      \caption{Hubble Type $T$ as a function of group-centric (\textit{left panel}) and cluster-centric distance (\textit{right panel}) for bright galaxies in the Fornax~A subgroup (\textit{left panel}) and Fornax cluster core (\textit{right panel}). Data for the cluster core galaxies are taken from \citet{Iodice2019} for ETGs and \citet{raj19} for LTGs.}
    \label{T_grp}
   \end{figure*}

The Fornax~A subgroup is populated with bright ($<16$ mag) LTGs and only one giant peculiar ETG, that is, NGC~1316 as its central galaxy. Most of the bright galaxies ($>75$ \%) in the sample are located in the north-east (NE) side of the Fornax~A group. In Fig.~\ref{T_grp} (left panel), we plot the morphological stage parameter $T$ as a function of projected distances from the group centre. This shows that the sample of LTGs is heterogeneous, similar to that of the LTGs in the Fornax cluster core (see right panel of Fig.~\ref{T_grp} and \citealt{raj19}). Overall, we do not find any morphological segregation of galaxies in the group.

\begin{figure*}[h]
%\hspace{-0.55cm}
% \vspace{-0.25cm}
   \includegraphics[width=9.4cm]{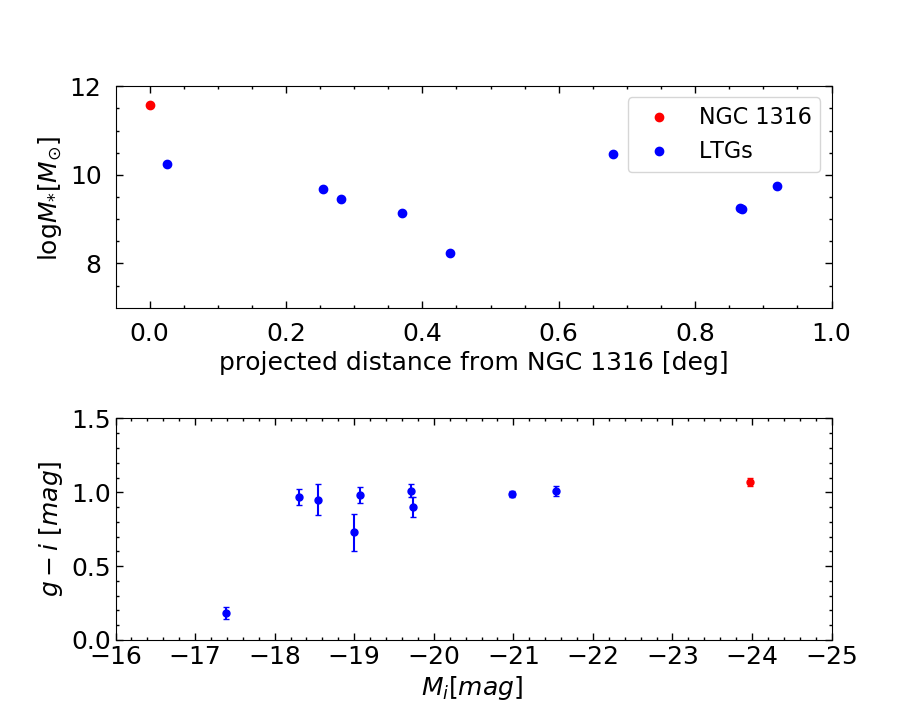}
    \includegraphics[width=9.45cm]{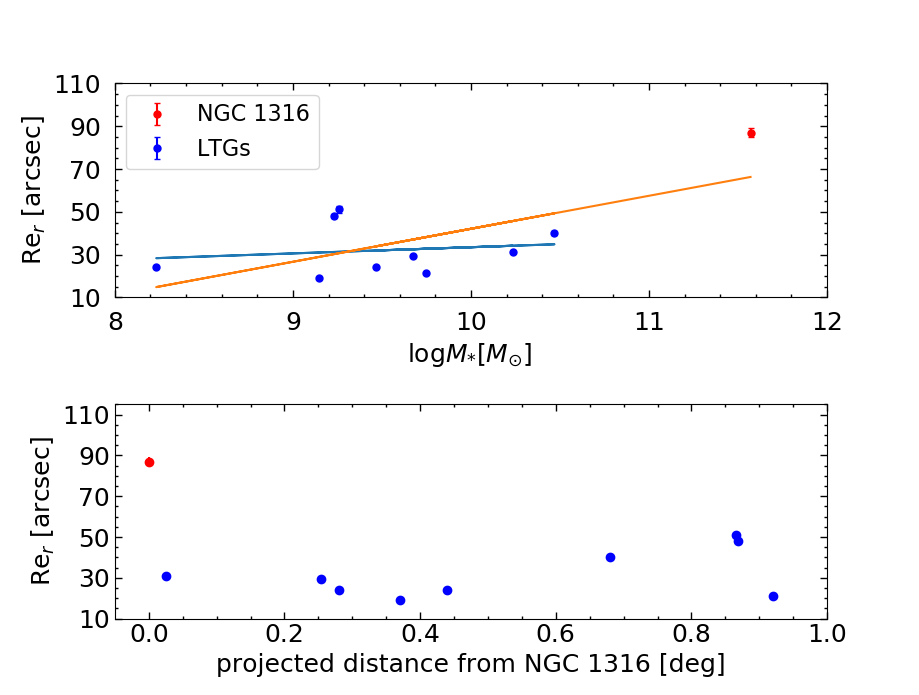}
      \caption{\textit{Left panels}: Stellar mass as function of projected distance from NGC~1316 (\textit{top panel}) and colour-magnitude relation (\textit{lower panel}) for bright spiral galaxies inside the virial radius of the Fornax~A sub group.
      \textit{Right panels}: Size-mass relation (\textit{top panel}) and effective radius as a function of projected distance from NGC~1316 (\textit{lower panel}) for all the bright spiral galaxies in the Fornax~A subgroup. A least square fit to the data is shown in orange (including NGC~1316) and blue (excluding NGC~1316).}
    \label{mass_grp}
   \end{figure*}

All LTGs in the Fornax~A group have stellar masses in the range  $8< \rm{log} \mathit{(M_*)[M_{\odot}}] < 10.5$, uniformly distributed inside the group (see top-left panel of Fig.~\ref{mass_grp}). As expected, from the colour-magnitude relation, the fainter galaxies ($M_i \leq -18$~mag) in the group have bluer colours, $g-i \leq 0.5$~mag (see lower left panel of Fig.~\ref{mass_grp}).

ETGs and LTGs follow a different size-mass relation \citep[see][]{Shen03} and this is likely the reason for the difference in the size-mass relation driven by the massive galaxy NGC~1316 and a less-pronounced relation observed for the bright spiral galaxies in the Fornax~A subgroup (Fig.~\ref{mass_grp}: top right panel). We do not find any correlation of effective radius as a function of the projected distance from NGC~1316 (Fig.~\ref{mass_grp}: lower right panel). The effective radii of all bright spiral galaxies in the Fornax~A subgroup are R$_e\leq$~70~arcsec ($\sim 7.07$ kpc).

\begin{figure}[h]
%\hspace{-0.55cm}
% \vspace{-0.25cm}
   \includegraphics[width=9cm]{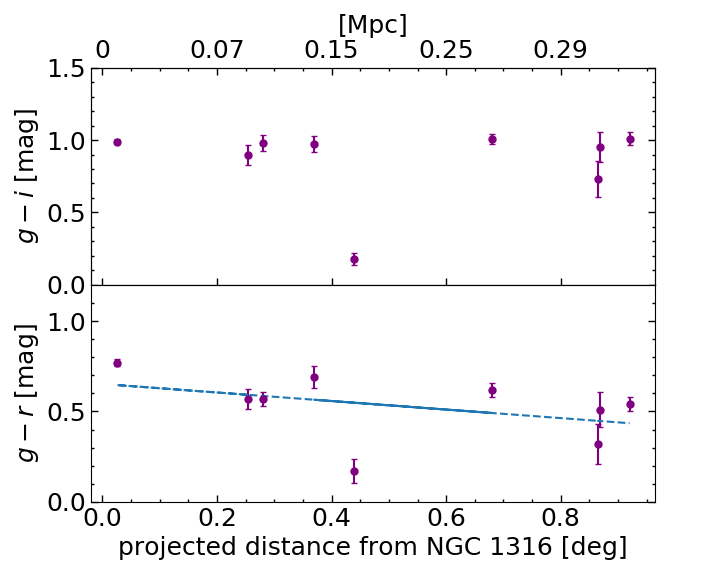}
      \caption{$g-i$  (\textit{top panel}) and $g-r$ (\textit{lower panel}) colours as a function of projected distance from NGC~1316 in degrees and Mpc. A least square fit to the $g-r$ data is shown. Slope of the fit is 0.23 $\pm$ 0.15.}
    \label{clr_grp}
   \end{figure}

As found for the other properties derived for the LTGs in Fornax~A group (i.e. morphology and stellar mass), the colour distribution also appears to be quite uniform inside the group, without any evident colour segregation with the group-centric distance (see Fig.~\ref{clr_grp}). A slight, but not significant trend (Pearson's correlation coefficient and the correlation significance are $-$0.41 and 0.26) might be observed in ($g-r$) colours as a function of projected distance, whereas, on average, most ($\sim$~90\%) of the LTGs in the Fornax~A subgroup have colours in the range ${0.7 \leq g-i \leq 1}$~mag. The very blue outlier FCC035 (${g-i = 0.17\pm0.1}$~mag) has a nuclear star cluster in its central region and this contributes to its blue ($g-i$) colour.

\subsection{Analysis of the discs}

Similar to our analysis in \citet{raj19} for the galaxies in the virial radius of the Fornax cluster, we study the discs of the galaxies in the group as well. Our results of the disc-breaks show that six galaxies have Type~III (up-bending) profiles. Even considering that this is a small sample, we examine the main correlations of the disc-breaks with other galaxy properties (e.g. stellar mass, total magnitude) inside the Fornax~A group. Results are also compared with those derived for the LTGs in the Fornax cluster core by \citet{raj19}.

\begin{figure*}[h]
 \centering
   \includegraphics[width=15cm]{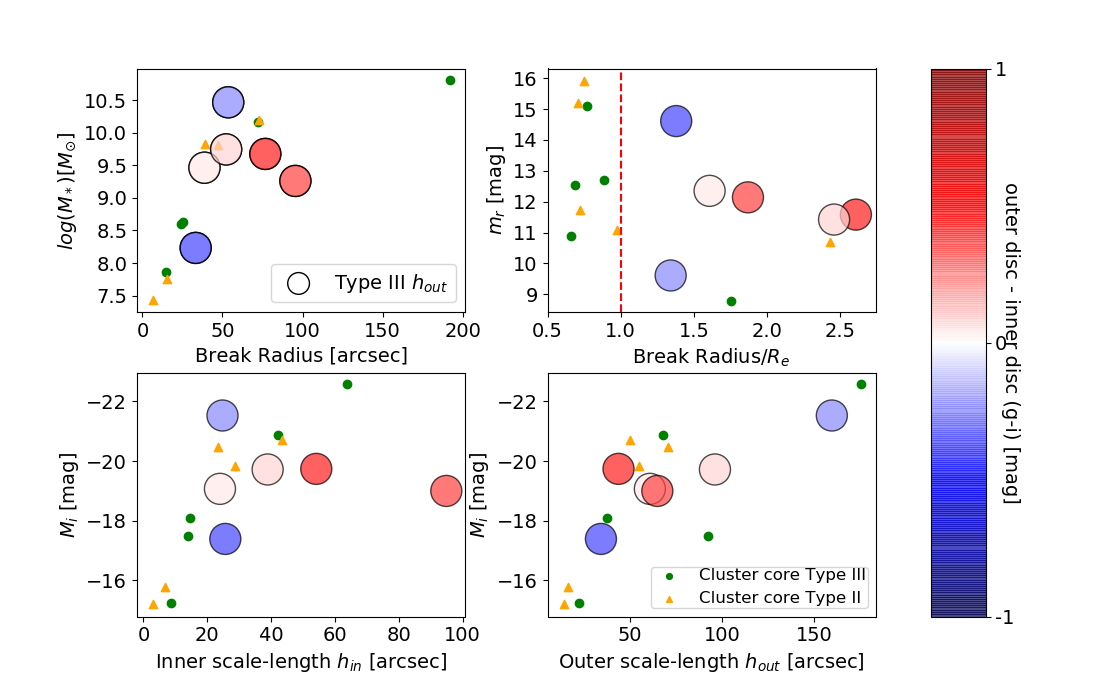}
      \caption{Analysis of galaxies with a break in their surface brightness profiles. Circles represent Type~III, with average ($g-i$) colour of $h_{\rm{out}}-h_{\rm{in}}$ as colour map. Orange triangles and green circles represent Type~II and Type~III galaxies in the Fornax cluster core \citep[data taken from][]{raj19}. \textit{Top panel left}: Stellar mass as a function of break radius. \textit{Top panel right}: Total magnitude as a function of break radius normalised to effective radius. The red dashed line indicates 80\% of the Fornax cluster core galaxies with disc-breaks. \textit{Bottom panels}: Absolute magnitude in $i$-band as a function of  inner scale-length $h_{\rm{in}}$ (\textit{left}) and outer scale-length $h_{\rm{out}}$ (\textit{right}).}
         \label{br_analysis}
         \end{figure*}

We do not find any disc-break trends with regard to the stellar mass (see top left panel of Fig.~\ref{br_analysis}). 
The same can be said for (i) total magnitude as a function of break radius normalised to the effective radius in $r$ (top right panel); (ii) absolute magnitude as a function of inner-scale length, $h_{\rm{in}}$ (lower left panel in Fig.~\ref{br_analysis}). A less-pronounced relation between the outer scale-lengths and absolute magnitude is observed for the LTGs in the Fornax~A subgroup (lower right panel in Fig.~\ref{br_analysis}). 
Galaxies in the Fornax~A group have disc-breaks located at different scale lengths, irrespective of their effective radius and total magnitude, unlike the galaxies in the cluster core 
\citep{raj19} where there were observed trends such as, increasing stellar mass with $B_r$ (see also cluster core LTGs in Fig.~\ref{br_analysis}). For all six galaxies in the Fornax~A group, the break radius is located beyond 1$R_e$. 

Except for NGC~1326 ($g-i_{h_{\rm{out}} - h_{\rm{in}}}=-0.24$ mag) and FCC~035 ($g-i_{h_{\rm{out}} - h_{\rm{in}}}=-0.36$~mag), the other four galaxies show redder outer discs. FCC~035 is a very late-type galaxy ($T=9$) and is the faintest galaxy with minor disturbances in its outskirts (see Fig.~\ref{FCC035}). NGC~1326 ($T=1$) is the brightest in the LTG sample (see Fig.~\ref{FCC029}) with a double bar \citep{Erwin05}.  

\section{Assembly history of the bright LTGs in the Fornax~A subgroup} \label{history}

In this section, we combine the main results for the brightest galaxies in the Fornax~A group (morphology and disc structure) in the two-dimensional projected phase-space (PPS) to trace its assembly history. Recently, studies concerning the location of galaxies in phase-space have gained importance \citep[e.g.][]{Mahajan2011, Hernan2014, Muzzin2014, Rhee2017}. These studies show the connection between a galaxy's phase-space location and its effect on star formation and stellar mass. This helps to further understand the effect of the environment on the evolution of galaxies, following on the morphology-density relation \citep{Dressler1980}. 

\begin{figure*}
%\hspace{-0.55cm}
% \vspace{-0.25cm}
  \centering
   \includegraphics[scale=0.22]{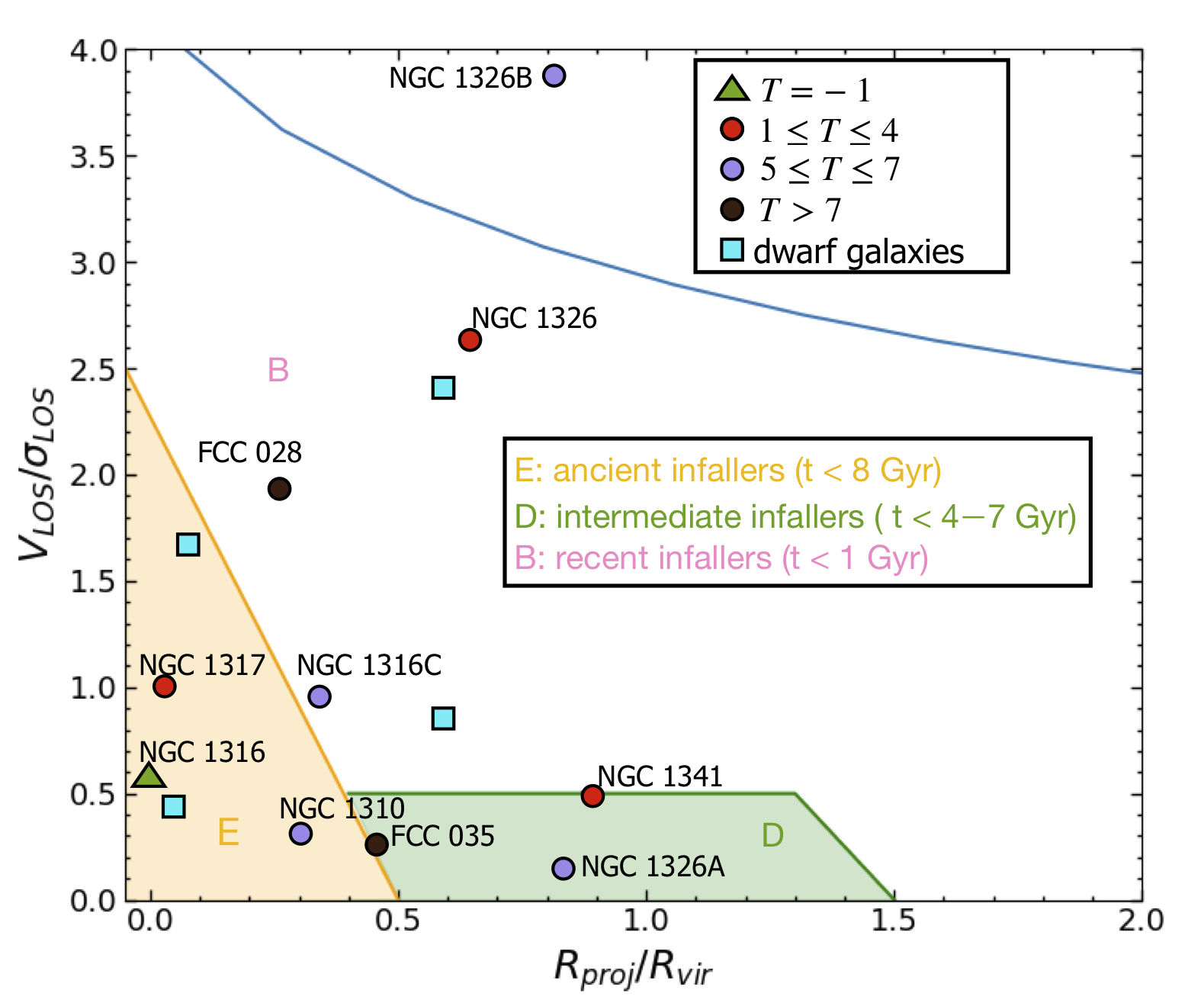}
      \caption{Projected phase-space diagram of galaxies in the Fornax~A subgroup within a group-centric distance of 1.05 deg $\sim$ 0.38 Mpc. The blue trajectory corresponds to the escape velocity of the group. Regions E (orange), D (green), B (pink) represent ancient, intermediate, and recent infallers as defined by \citet{Rhee2017}. The bright spiral galaxies are indicated with coloured circles based on their morphological type $T$ (red, purple, brown), NGC~1316 is indicated with a green triangle ($T=-1$) and dwarf galaxies are indicated with blue squares.}
    \label{pps}
   \end{figure*}

We used the published catalogue of spectroscopic redshifts by \citet{Maddox2019} to plot the phase-space diagram for 15 galaxies part of the Fornax~A subgroup, including dwarfs. This is shown in Fig.~\ref{pps}. \citet{Maddox2019} use a sample of 40 galaxies within a radius of 0.7 Mpc and find the central velocity of the Fornax~A subgroup to be 1778 km s$^{-1}$ with a velocity dispersion of 204 km s$^{-1}$, and estimate the dynamical mass of the Fornax~A subgroup to be 1.6 $\times 10^{13} M_{\odot}$. According to \citet{Maddox2019}, we adopted 1.05 deg ($\sim$ 0.38 Mpc) as the virial radius of the Fornax~A subgroup.  

In Fig.~\ref{pps}, we plot the velocities (normalised to the group velocity dispersion) as a function of the projected distance (normalised to the virial radius). Along with this, we also plot the trajectory curve for the escape velocity\footnote{ The escape velocity $v_{esc}$ =$ \sqrt{\frac{2GM_{vir}}{R_{vir}}K(s)}$, where $K(s)= g_c \frac{ln(1+cs)}{s}$, $s=\frac{R_{3D}}{R_{proj}}$ and $g_c=[ln(1+c) - \frac{c}{1+c}]^{-1}$.} ($v_{esc})$. According to \citet{Rhee2017}, in the PPS we have also included the regions populated by the galaxies that merged into the cluster potential at different infalling times. They are the "ancient infallers" ($t < 8 $~Gyr), "intermediate infallers" ($t < 4\text{--}7$~Gyr), and "recent infallers" ($t < 1$~Gyr). \citet{Rhee2017} explain the origin of these regions by comparing them to the trajectory such that galaxies falling for the first time are close to the escape velocity curve, and as they pass through the pericentre, they start gaining velocity. Over time, after passing through the apocentre and many more pericentres, they eventually end up in the region of ancient infallers. 

The PPS for the Fornax~A group shows that the majority of the group members are located in the region of the recent and intermediate infallers. This would suggest that most galaxies are still assembling into the Fornax~A group. The ancient infallers are the central and massive galaxy NGC~1316 and the close (in projection) barred spiral galaxy  NGC~1317 (see also Fig.~\ref{FCC022}). Most of the bright spiral galaxies in the Fornax~A group (NGC~1310, NGC~1326A, FCC~035, and NGC~1341) seem to be intermediate infallers. Except for FCC035, galaxies in this region have average colour ${0.9 < g-i< 1}$~mag. Although the Fornax~A group is not as dense as the Fornax cluster, the group potential should somehow impact the outskirts of the discs. This is what we mainly observe for NGC~1316C, FCC~028, NGC~1326 and NGC~1326B (see Fig.~\ref{FCC028}, Fig.~\ref{FCC033}, Fig.~\ref{FCC037} and Fig.~\ref{FCC039}) that belong to the region of recent infallers. NGC~1326, the most massive LTG in the sample, also shows plenty of star-forming regions.  

Six of the galaxies in the region of intermediate and recent infallers also show a Type~III disc-break. 
We explain the possible mechanisms for the formation of this break in Sect.~\ref{core_grp}.  Finally, according to the PPS, the interacting system NGC~1326A and NGC~1326B may not be interacting, and that NGC~1326B is experiencing a fair amount of tidal interaction through its first infall, which would explain the star-forming regions in its outer-disc. 

%-----------------------------------------------------------------------------------------

\section{The build-up of the IGL in the Fornax~A subgroup}\label{igl_disc}

In this section, we summarise relevant results from this work and those from previous works on the Fornax~A group, which help to understand the formation history of its IGL. We then discuss how predictions from models fit in well with our results from observations.

\begin{figure*}[h]
   \centering
%\hspace{-0.55cm}
% \vspace{-0.25cm}
   \includegraphics[scale=0.25]{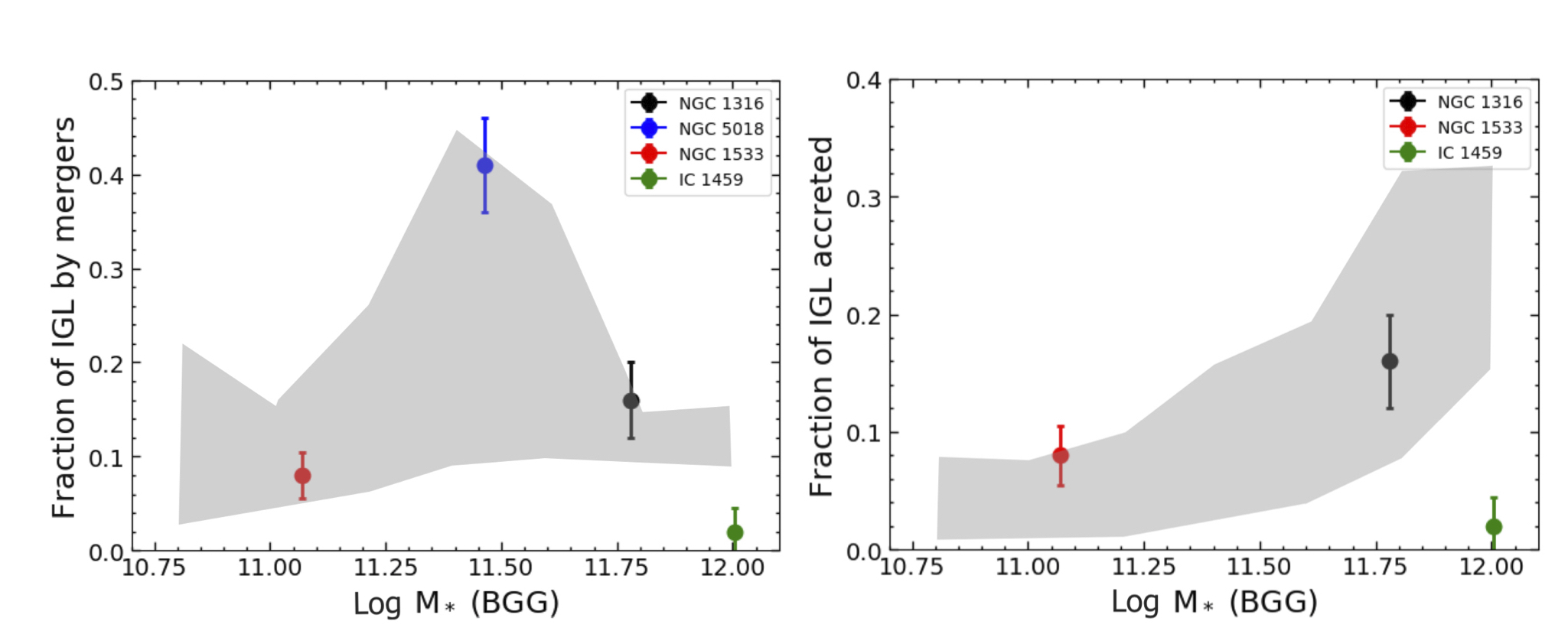}
      \caption{Fraction of IGL by mergers (\textit{left panel}) and fraction of IGL accreted (\textit{right panel}), as a function of the BGG stellar mass for four groups: NGC~1316, NGC~5018 \citep{spavone18}, NGC~1533 \citep{cattapan19}, IC~1459 \citep{Iodice2020}. The shaded areas represent the 20$^{\rm{th}}$ and 80$^{\rm{th}}$ percentile of the "model disruption" described by \citet{Contini14}.}
    \label{igl_grp2}
   \end{figure*}

The IGL in the Fornax~A group is estimated within a region $\sim$ 35 arcmin$^2$ ($\sim$0.5~deg), which covers most of the group extension (see Sect.~\ref{sec:IGL_method}). 
Inside this radius, \citet{venhola2019} find a significant drop in the number of dwarf galaxies ($\sim$ 50\%).
This might be an indication that the IGL in the Fornax~A group was assembled by past merger of NGC~1316 \citep[see][]{Iodice2017,serra19} and disruption of satellite galaxies on infall, which also contributed to the build-up of the extended stellar envelope around NGC~1316. This hypothesis is further supported by comparing the IGL fraction in the Fornax~A subgroup with theoretical predictions from \citet{Contini14} (see Fig.~\ref{igl_grp2}). 
According to \citet{Contini14}, galaxies which get destroyed while falling into the cluster or group make the accreted component of the ICL or IGL.  
Since the IGL in the Fornax~A subgroup fits well in both regions from tidal disruption and merger channels, this suggests that at the present epoch, both mechanisms contributed to the build-up of this component.

In Fig.~\ref{igl_grp2} we also compare the IGL derived for the Fornax~A group with values obtained for other groups of galaxies, from observations of comparable depth and with the same image analysis. In particular, we plot the fraction of IGL to the total luminosity of the bright member galaxies as a function of the BGG stellar mass (in the range $10^{11}$ -- $10^{12}$~$M_{\odot}$) for four groups: Fornax~A (NGC~1316), 
NGC~5018 \citep{spavone18}, NGC~1533 \citep{cattapan19}, and IC~1459 \citep{Iodice2020}. Compared with theoretical predictions, the IGL fraction in each group reflects their assembly history. The main contribution to the IGL in the NGC~5018 group comes from the extended tidal tail that is tracing an ongoing merger \citep{spavone18}. 
In the Fornax~A group, as well as in the NGC~1533 triplet \citep{cattapan19}, 
that have a comparable IGL fraction, mergers would have partly contributed to the IGL, and signs of past mergers are confined in the BGG's envelope (in the form of loops and shells). As pointed out by \citet{Iodice2020}, given the low IGL fraction ($\sim 2\%$) and the structure and properties of galaxy members, the IC~1459 group is still in an early 
mass assembly phase.  

The connection between the IGL fraction and evolution is also consistent with the different ratios of ETG-to-LTG galaxies. In the Fornax~A group, ETG/LTG$=$0.11, since it has only one peculiar ETG (NGC~1316) and nine bright LTGs. NGC~5018 and NGC~1533 groups have ETG/LTG$=1.6$, while IC~1549 ETGs/LTGs$=0.3$. Although the Fornax~A subgroup has the lowest ratio among the groups discussed above, a major fraction of the IGL with respect to the IC~1459 group, which has comparable ETG-to-LTG ratio, comes from the past merger of NGC~1316 and disrupted dwarf galaxies. In the IC~1459, no signs of interactions are found with a galaxy of comparable mass \citep{Iodice2020}.  

%-----------------------------------------------------------------------------------------

\section{The tale of the Fornax environment: The cluster core versus the infalling group } \label{core_grp}

It is known that galaxies are redder and brighter in dense environments than in low-density environments \citep{faber1973,oemler1974,visvanath77, butcher84}. The Fornax environment, comprising the cluster core and the Fornax~A group, can depict "pre-processing" of galaxies, in particular, how LTGs evolve to the observed ETGs in the cluster core. From literature, the proposed physical processes that galaxies undergo in groups and subclusters are strangulation and merging. In this section, we compare the properties of the group to that 
of the cluster core thus showing the different evolutionary stages for these environments. 

Galaxy populations: The Fornax cluster core has 18 bright (${m_B< 15}$ mag) ETGs \citep{Iodice2019} and 13 bright (${m_B< 16.6}$ mag) LTGs \citep{raj19}. The total number of dwarf galaxies in the cluster is 564 \citep{Venhola2018, venhola2019}. However, the Fornax~A group has nine bright ($m_B<$ 16 mag) LTGs and one peculiar ETG, that is the central galaxy NGC~1316. The total number of dwarf galaxies is $\sim$ 60 within $R_{vir}=1.0$~deg \citep{venhola2019}.  Overall, the number of galaxies in the Fornax cluster core is nearly ten times more than that of the Fornax~A group.

Morphological segregation: Nearly 70\% of the ETGs (${T\leq-2}$) are located within 0.8 deg of the cluster core (see Fig.~\ref{T_grp}). 
This region, where the bulk of the X-ray emission is also found, correspond to the high-density region of the cluster \citep{Iodice2019}. Beyond this region, that is $D_{core}>$ 1 deg, 60\% of the LTGs are located \citep[see][]{raj19}. 
These results are consistent with the morphology-density relation \citep{Dressler1980}. We do not observe the same for the galaxies in the Fornax~A group (see Fig.~\ref{T_grp}). This suggests that the environment is not as dense as the Fornax cluster core, where the late-type star-forming galaxies are slowly transforming into passively evolved cluster members \citep[e.g.][]{oemler1974}. The absence of ETGs in the Fornax~A group and the morphological segregation confirm that the group is at a different stage of evolution. 
   
\begin{figure*}[h]
%\hspace{-0.55cm}
% \vspace{-0.25cm}
\centering
   \includegraphics[width=18cm]{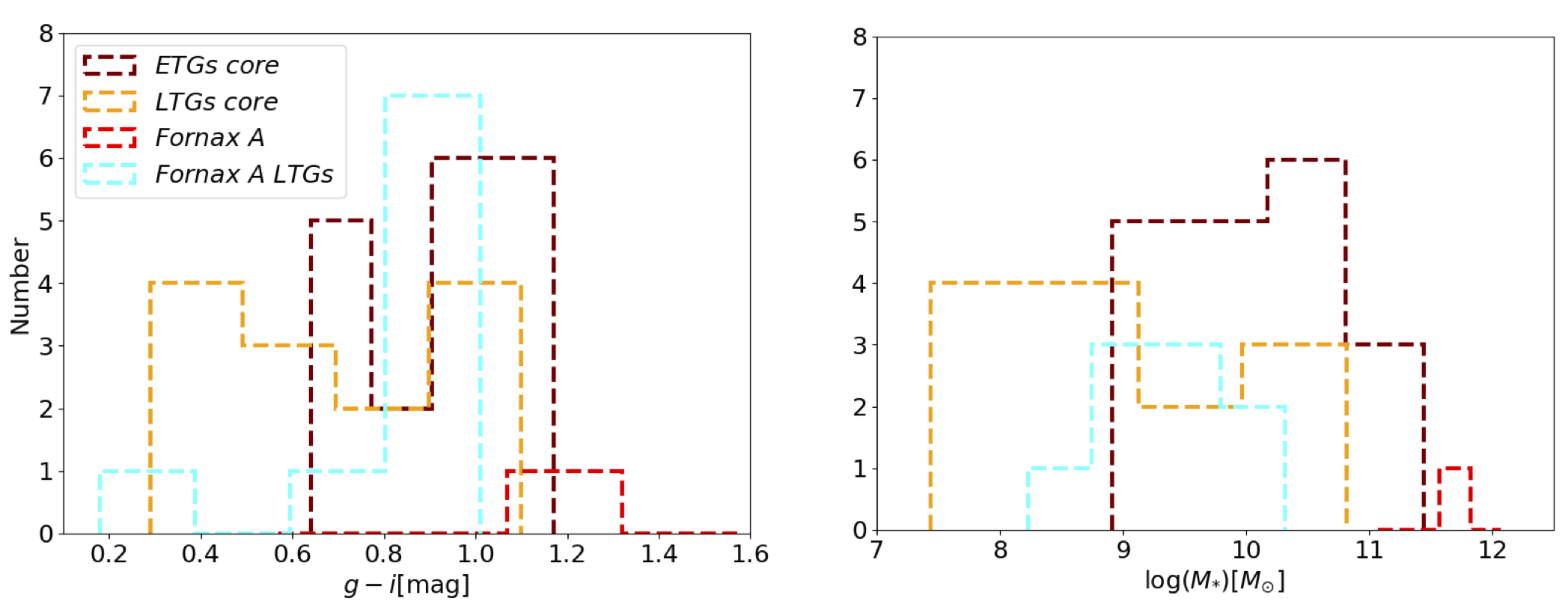}
     \caption{Histogram of the colour ($g-i$) distribution (\textit{left panel}) and of the stellar mass (\textit{right panel}) for the bright galaxies in the Fornax core and Fornax~A subgroup. }
    \label{fig:col_mass_hist}
   \end{figure*}

Colour segregation: For galaxies in the cluster core, a strong colour segregation as a function of the cluster-centric radius is observed, where the redder and massive ETGs populate the high-density region of the cluster \citep{Iodice2019}. Most of the blue LTGs are found in the low-density region of the cluster. The Fornax~A group, dominated by LTGs, does not show any colour segregation (see Fig.~\ref{clr_grp}). The LTGs are about the same in ($g-i$) colours  as the LTGs in Fornax core (see left panel of Fig.~\ref{fig:col_mass_hist}). Figure~\ref{fig:col_mass_hist} (right panel) also shows that there are LTGs with lower stellar mass in the Fornax core ($7<\rm{log}\mathit{(M_*)} [\mathit{M_{\odot}}] < 8$). Hence the overall stellar mass range of LTGs, 
both in the core and the group is ${7<\rm{log}\mathit{(M_*)} [\mathit{M_{\odot}}]<11}$.

Structure of LTGs: The LTGs in the cluster core exhibit very different substructures that are well-correlated with their locations in the cluster core \citep[see][]{raj19}. As for the LTGs in Fornax~A group, their substructures 
are almost similar showing regular star-forming regions in their spiral arms. Although we find that Hubble type $T$ is heterogeneous in both samples, LTGs in the core show more irregularities in their structures 
(e.g. lopsidedness, tidal tails, boxy discs) in comparison to those in Fornax~A group, where almost none 
show any such irregularities (except for NGC~1327B). This suggests that the environment of the cluster core 
has played a major role in the structure-transformation of LTGs, while galaxies in the Fornax~A subgroup are undergoing minor changes or have not undergone any changes yet.  

Disc-breaks of LTGs: We find that many of the galaxies in the Fornax core have undergone pre-processing, star formation quenching in their outskirts during their time in the cluster \citep{Zabel2018, raj19, iodice2019b}. This 
 is also clearly seen through their disc-breaks, that is, the break radius of most (80\%) of these galaxies is located within $1R_e$. Beyond this radius, there are no substructures (e.g. spiral arms) or star-forming 
 regions.  However, most of the disc-breaks seen in the group are of Type~III, in contrast to those in the cluster core. Moreover, the break radius of galaxies in the Fornax~A subgroup is located 
 beyond $1 R_e$. This suggests that the environmental processes in the group have not severely altered the structures of the galaxies therein. 
 
 Mergers and strangulation are common phenomena in the group environment \citep{Barnes85, Zablu98, hashimoto2000}, but this assumption is not enough to understand the origin of Type~III disc-breaks alone. 
 While they have been many explanations for their occurrence \citep[e.g.][]{Bakos2008, Younger2007, Laine2016, watkins19}, we state a few which depend on the H\textsc{i} content, such that if the gas is not replenished, star formation is eventually halted.  It is also important to consider internal mechanisms such as the formation of bars, which can also form a Type~III disc-break \citep[e.g.][]{Erwin05, PohlenTrujillo2006}.\\ 
  \citet{Wang18} showed that Type~III disc-breaks were prevalent in H\textsc{i}-rich galaxies with low-spin. The \textit{in-situ} star formation from gas accretion builds the outer-disc of these Type~III galaxies. Alternatively, enhanced star formation in the inner-discs of galaxies can cause steeper inner-discs than outer discs \citep{Hunter06}. Gas in outer-discs are vulnerable to the tidal forces in a dense environment. In this case, the gas that falls inwards can cause star-bursts in the inner central regions \citep[e.g.][and references therein]{barnes91, barnes96, Hopkins13}. If LTGs in the Fornax~A group are undergoing strangulation, their outer-discs would not have cold gas to fuel star formation. H\textsc{i} data are required to confirm if Type~III disc-breaks in LTGs part of the Fornax~A group, have formed as a consequence of H\textsc{i}-richness or the contrary. 
  
  X-ray emission: The main contributions to the X-ray emission in Fornax~A group are provided by the centre of the BGG, NGC~1316, and by two of the most luminous stellar loops in the outskirts \citep[L1 in the south and L2 on the east of the galaxy,][]{Mackie1998}. In addition, there is an elongated region of X-ray emission to the NW of NGC~1316, corresponding to the region of the plume \citep{Iodice2017}. The total luminosity of the 
  X-ray emission is $3.1 \times 10^{40}$~erg s$^{-1}$ \citep{kim03} is an order of magnitude fainter than the total X-ray luminosity in the Fornax core which is $5 \times 10^{41}$~erg s$^{-1}$ \citep{Jones1997}. Moreover,
  the X-ray emission in the core of the cluster, which peaks on the BCG NGC~1399 is much more extended, reaching $\sim 0.4 R_{vir} \sim 0.3$~Mpc. Most of the brightest and massive galaxies in the core are located inside the X-ray emission and all of them are ETGs \citep{Iodice2019}. Therefore, most of the processes linked to the galaxy harassment that happened in this region of the cluster, and that shaped most of the observed galaxy properties, cannot be taken into account in the Fornax~A group. This is further consistent with the homogeneous population of LTGs found in this group and with the overall picture that the Fornax~A group is not as evolved as the cluster core.

\section{Summary and conclusions} \label{conclu3}

With FDS data, we have analysed the structure of the brightest ($m_B< 16$~mag) LTGs in Fornax~A SW group of the Fornax cluster. We have also taken advantage of the depth of the data and derived the IGL amount in the Fornax~A group. In the study of disc-breaks of LTGs in different environments, we have compared the break radius of LTGs in the cluster core to those of the group. Since the Fornax cluster has been studied by several surveys, we have made the best use of available public data, in this case, the spectroscopic redshift compilation by \citet{Maddox2019} to look at the group members in phase-space. Thus, for the first time, we give a complete picture of the environment of the Fornax~A group, and compare it to that of the Fornax cluster core. 
We summarise the results of this work as follows: 
\begin{enumerate}

    \item In contrast to what is observed in the core of the Fornax cluster, in the Fornax~A group, there is no observed morphological, mass and colour segregation of galaxies. LTGs in the 
    Fornax~A group have stellar masses in the range $8< \rm{log}(\mathit{M_*}) [\mathit{M_{\odot}}] <$ 10.5), while there are less massive LTGs in the Fornax cluster core ($7~<~\rm{log} (\mathit{M_*}) [\mathit{M_{\odot}}]<11$). 
    \item We detect only Type~III disc-breaks in the SB profiles of galaxies in the Fornax~A group. Of them, four galaxies have outer discs redder than their inner discs, and two galaxies have bluer outer discs. On the contrary, in the Fornax core, we detected all types of disc-breaks, suggesting a variety of mechanisms and interactions with the environment \citep{raj19}. 
    The formation of the breaks in the Fornax~A group needs more investigation of their H\textsc{i} content, in order to investigate whether H\textsc{i}-rich galaxies can cause enhanced star formation in the galaxy's outer-disc, or whether these galaxies are experiencing strangulation which halts star formation in the outer-discs.  
    \item The break radii of six galaxies in the Fornax~A subgroup are located beyond 1$R_e$. This is contrary to the results of the galaxies in the Fornax cluster core where the break radii of galaxies (80\% of the sample) are located within $1R_e$. This suggests that the environmental mechanisms in the Fornax~A group are not as strong as those in the cluster core, where the latter has altered the structures of the galaxies' disc. 
    \item Through phase-space analysis, we find that most (80\%) of the group members are 
    intermediate and recent infallers, consistent with a younger assembly epoch with respect to the Fornax core, where the PPS is equally populated by ancient infallers (that dominate the high-density region of the cluster) and by intermediate and recent infallers that are symmetrically distributed around the cluster core in the low-density region \citep{iodice2019b}.
    \item The luminosity of the IGL in the Fornax~A group is estimated to be $6\pm2\times 10^{10} L_{\odot} $ in $g$. Compared to the total luminosity of the group, the fraction of the IGL is $\sim 16\%$.
    Our estimates agree with model predictions from \citet{Contini14}. Thus, we address the formation history of the IGL in the Fornax~A group which is mainly through the past merger of NGC~1316 and the disruption of dwarf galaxies in the same region. This is further supported by the significant drop in the number of dwarf galaxies ($\sim 50\%$) towards the group centre found by \citet{venhola2019}.
   
\end{enumerate}

According to the above results and the whole analysis performed in this work, the Fornax~A group appears to be in an early-stage of assembly with respect to the cluster core.
The environment of the Fornax~A subgroup is not as dense as that of the cluster core, with all galaxies except the BGG showing similar morphology, comparable colours and stellar masses, and Type~III disc-breaks, without any clear trend of these properties with group-centric distances. The low amount of IGL is also consistent with this picture, since there were no significant gravitational interactions between galaxies that modified the galaxy structure (forming tidal tails or stellar streams) and contributed to the build-up of the IGL. 
The main contribution to the IGL is from the minor merging in the outskirts of the BGG NGC~1316 and, probably, the disrupted dwarf galaxies close to the group centre.

As future perspectives involve the analysis of the star formation and, therefore, the comparison of it with the detected disc-breaks, we will study the H$\alpha$ distribution of the brightest LTGs in the Fornax~A group (Raj et al. in prep). A comparison of the overall structural parameters of galaxies in the cluster core and the group is currently analysed by Su et al. (in prep). All of the mentioned work is part of FDS, thus providing a detailed analysis of the Fornax environment and galaxies therein. 

\begin{acknowledgements}
The authors are grateful to the anonymous referee for the useful comments and suggestions that improved this article. This publication has received funding from the European Union Horizon 2020 research and innovation programme under the Marie Sk\l odowska-Curie grant agreement n. 721463 to the SUNDIAL ITN network. This work is based on visitor mode observations collected
at the European Organisation for Astronomical Research in the Southern Hemisphere under the following VST GTO programs: 094.B-0512(B), 094.B-0496(A), 096.B-0501(B), 096.B-0582(A). NRN acknowledges financial support from the “One hundred top talent program of Sun Yat-sen University” grant N. 71000-18841229. J.F-B acknowledges support through the RAVET project by the grant AYA2016-77237-C3-1-P from the Spanish Ministry of Science, Innovation and Universities (MCIU) and through the IAC project TRACES which is partially supported through the state budget and the regional budget of the Consejer\'\i a de Econom\'\i a, Industria, Comercio y Conocimiento of the Canary Islands Autonomous Community. GvdV acknowledges funding from the European Research Council (ERC) under the European Union's Horizon 2020 research and innovation programme under grant agreement No 724857 (Consolidator Grant ArcheoDyn). This research has made use of the NASA/IPAC Extragalactic Database (NED), which is operated by the Jet Propulsion Laboratory, California Institute of Technology,
under contract with the National Aeronautics and Space Administration. We also thank Dr. Johan Knapen for his suggestions and comments. 
\end{acknowledgements}
\bibliographystyle{aa.bst} % style aa.bst
  \bibliography{fornaxa} 
  \begin{appendix}
  
  \section{Bright LTGs in the Fornax~A subgroup} \label{ap_A}
  In this section, we briefly describe the galaxies analysed in this work. 
  
  \subsection{FCC013}
  Also known as NGC~1310, is a barred-spiral galaxy (SBcII) with morphological type $T=5$, located at a projected distance $D_{core}= 0.25$ deg from NGC~1316. The H\textsc{i} disc of this galaxy is more extended than its stellar disc, with $M{_{\textsc{hi}}= 48.0\pm0.9 \times 10^7 M_{\odot}}$ \citep{serra19}. The galaxy has an effective radius ${R_{e,r} = 27.6}^{\prime\prime}$ and star-forming spiral arms. We detect a Type~III disc-break in its SB profile and, hence, the break radius is 7.8 kpc. The outer disc appears redder than the inner disc with average ${(g-i)_{h_{\rm{out}}-h_{\rm{in}}}= 0.44}$~mag. The origin of the Type~III break in this galaxy is likely due to its high H\textsc{i} content.  
  \subsection{FCC022}
\begin{figure*}
  \centering
   \includegraphics[width=\hsize]{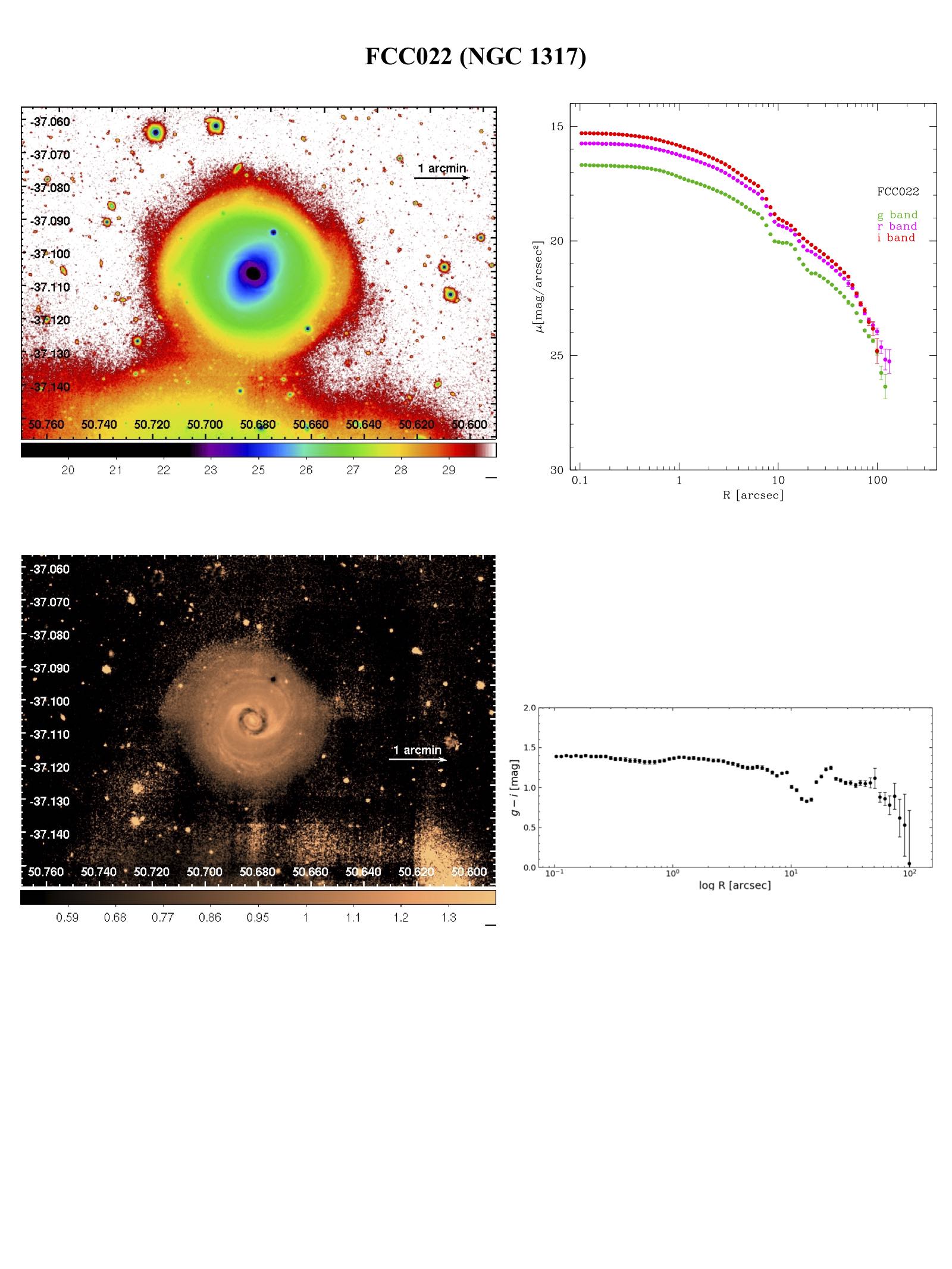}
     \vspace{-6.5cm}
     
      \caption{Surface Photometry of FCC022. \textit{Top left panel}: $g$-band image in SB level. \textit{Top right panel}: SB profiles in $g, r, i$ bands. \textit{Bottom panels}: $g-i$ colour map (\textit{left}) and $g-i$ colour profile (\textit{right}).}
               
         \label{FCC022}
   \end{figure*}

  Also known as NGC~1317, is a spiral galaxy (SA pec) with morphological type $T=1$, located at closest projected distance from NGC~1316, that is, $D_{core}= 0.025$ deg from NGC~1316. The H\textsc{i} disc of the galaxy is confined to its stellar disc and has H\textsc{i} mass $M_{\textsc{hi}}= 26.4\pm0.6 \times 10^7 M_{\odot}$ \citep{horell01, serra19}, while in the central part of this galaxy, that is, at $R\sim$ 0.5$^{\prime}$, a star-forming ring is visible. It appears to be an early-type spiral, with almost no star-forming regions beyond R$\sim$ 0.5$^{\prime}$, which can be seen in its SB image (Appendix \ref{FCC022}). This can be justified with the results from phase-space analysis that FCC022 has passed through the pericentre and lost its neutral gas through interactions with the intra-group medium \citep[see also][]{serra19}. 
  \subsection{FCC028} 
  FCC028 is a spiral galaxy (SmIII) with morphological type $T=9$, located at projected distance $D_{core} = 0.28$ deg from NGC~1316. From literature, the galaxy has been associated to a ring galaxy \citep{jane86, corwin85, lau82}. The ring can be seen as star forming regions in the SB image \ref{FCC028}, which has likely originated from a collision \citep{Maddox2019}. The H\textsc{i} disc of the galaxy is more extended than its stellar disc and has H\textsc{i} mass $M_{\textsc{hi}}= 13.7\pm0.6 \times 10^7 M_{\odot}$  \citep{serra19}. We detect a Type~III disc-break in its SB profile and its break radius is at 3.9 kpc. The outer disc of this galaxy is redder than its inner disc with an average ${(g-i)_{h_{\rm{out}}-h_{\rm{in}}}=0.04}$~mag.  
  \subsection{FCC029}
  Also known as NGC~1326, is a barred-spiral galaxy (SBa(r)) with morphological type $T=1$ is located at a projected distance $D_{core}= 0.68$ deg from NGC~1316. \citet{peter04} describe it as a double barred galaxy. It is the most luminous, massive galaxy in our sample of LTGs with $M_i= -21.53$ and $M_*= 2.94 \times 10^{10} M_{\odot}$. We detect a Type~III disc-break with a break radius = 5.4 kpc. The outer disc is bluer than its inner disc with $(g-i)_{h_{\rm{out}}-h_{\rm{in}}}=-0.24$~mag. This can be associated to the star-forming regions in its outskirts. 
  \subsection{FCC033}
   Also known as NGC~1316C, is a spiral galaxy (SdIII pec) with morphological type $T=7$, located at a projected distance ${D_{core}= 0.37}$~deg from NGC~1316. The spiral arms are within $R=10^{\prime\prime}$. The effective radius of this galaxy is ${R_{e,g}=22.6 \pm 0.5}^{\prime\prime}$ and has a faint corona \citep{corwin85}.   
  \subsection{FCC035}
  FCC~035 is a spiral galaxy (SmIV) with morphological type ${T=9}$, and is located at a projected distance ${D_{core}= 0.44}$~deg from NGC~1316. It is the faintest galaxy in our sample ${M_i= -17.39}$ and ${M_*= 0.017 \times 10^{10} M_{\odot}}$. We detect a Type~III disc break in its SB profile with break radius ${= 2.8}$ kpc. The outer disc is bluer than its inner disc with ${(g-i)_{h_{out}-h_{in}}= -0.36}$~mag. The H$\alpha$ (Raj et al. in prep) distribution in the galaxy is concentrated in a star-forming region located in its spiral arms, which explains the blue ($g-i$) colours. This can also be associated to its very-late morphological type and irregularities in its spiral structure. 
  \subsection{FCC037} \label{037a}
  Also known as NGC~1326A, is a barred-spiral galaxy (SBcIII) with morphological type $T=5$, located at a projected distance ${D_{core}=0.87}$~deg. This galaxy has been studied as an interacting system along with FCC~039 (NGC~1327B). However, from our analysis in PPS (data taken from \citealt{Maddox2019}), we find the velocity difference between these galaxies is large and thus, they might not be interacting. 
  
  \subsection{FCC039}
  Also known as NGC~1327B, is a spiral galaxy (SdIII) with morphological type $T=7$, located at a projected distance ${D_{core}=0.86}$~deg from NGC~1316. This galaxy has irregular spiral (dusty) arms, and disruptions in outer disc as a result of its recent interaction with the tidal field of the Fornax environment. We detect a Type~III disc-break in its surface brightness profile and a break radius ${= 9.6}$~kpc. As mentioned in Appendix \ref{037a}, this galaxy might not be interacting with FCC~037 (NGC~1327A). However, it can be a fly-by thus showing faint tidal tails in its outskirts. The outer disc of this galaxy is redder than its inner disc with $(g-i)_{h_{\rm{out}}-h_{\rm{in}}}$= 0.36 mag.
  
  \subsection{FCC062}
  Also known as NGC~1341, is a barred-spiral galaxy (SbcII) with morphological type $T = 4$, located at furthest distance from NGC~1316, that is, ${D_{core}=0.92}$~deg. We detect a Type~III disc-break in its SB profile with disc break ${= 5.28}$~kpc. The outer disc of this galaxy is redder than its inner disc with spiral arms, Type~III disc break with ${(g-i)_{h_{\rm{out}}-h_{\rm{in}}}= 0.08}$~mag. 
 \clearpage 
 \onecolumn
 \begin{multicols}{2}
  \section{The Sample: Images and profiles}\label{sb_im}
In this section, we show the $g$-band VST images (in surface brightness level) of LTGs in the Fornax~A subgroup, surface brightness profiles, $g-i$ colour maps and ($g-i$) profiles. For six of the galaxies with disc-breaks, we mark the break radius on the $g$-band images in surface brightness level (black dotted lines), the $g-i$ colour maps (white dotted lines), and ($g-i$) colour profiles (red dotted lines).
 \end{multicols}
 
 \begin{figure*}[!h]
  \centering
   \includegraphics[width=\hsize]{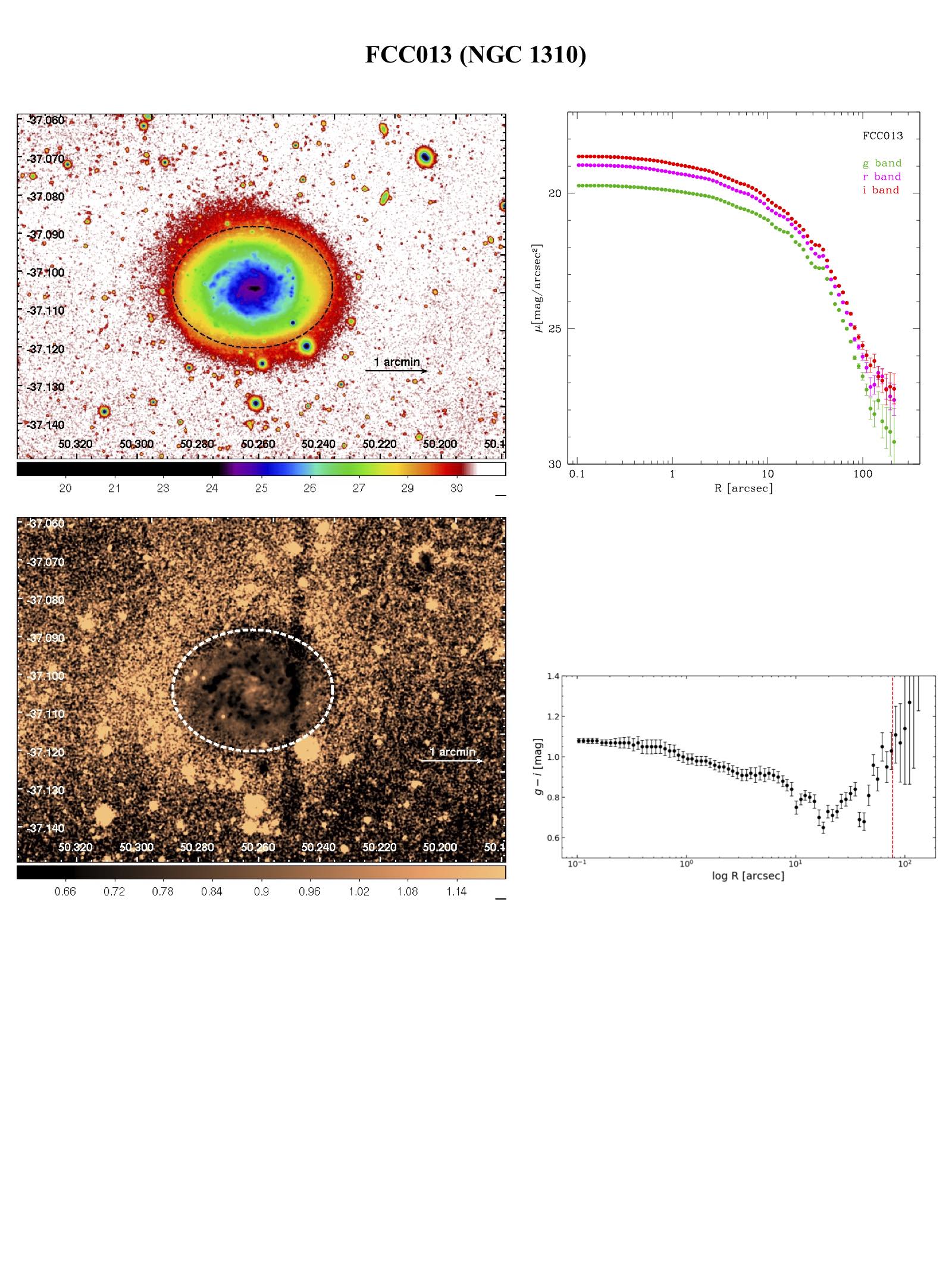}
     \vspace{-6.5cm}
       \caption{Surface Photometry of FCC013. \textit{Top left panel}: $g$-band image in SB level with black dotted lines representing the break-radius. \textit{Top right panel}: SB profiles in $g, r, i$ bands. \textit{Bottom left panel}: $g-i$ colour map with white dotted lines indicating the break-radius. \textit{Bottom right panel}: $g-i$ colour profile with red dotted line indicating the break radius.}
               
         \label{FCC013}
   \end{figure*}
 
\begin{figure*}
  \centering
   \includegraphics[width=\hsize]{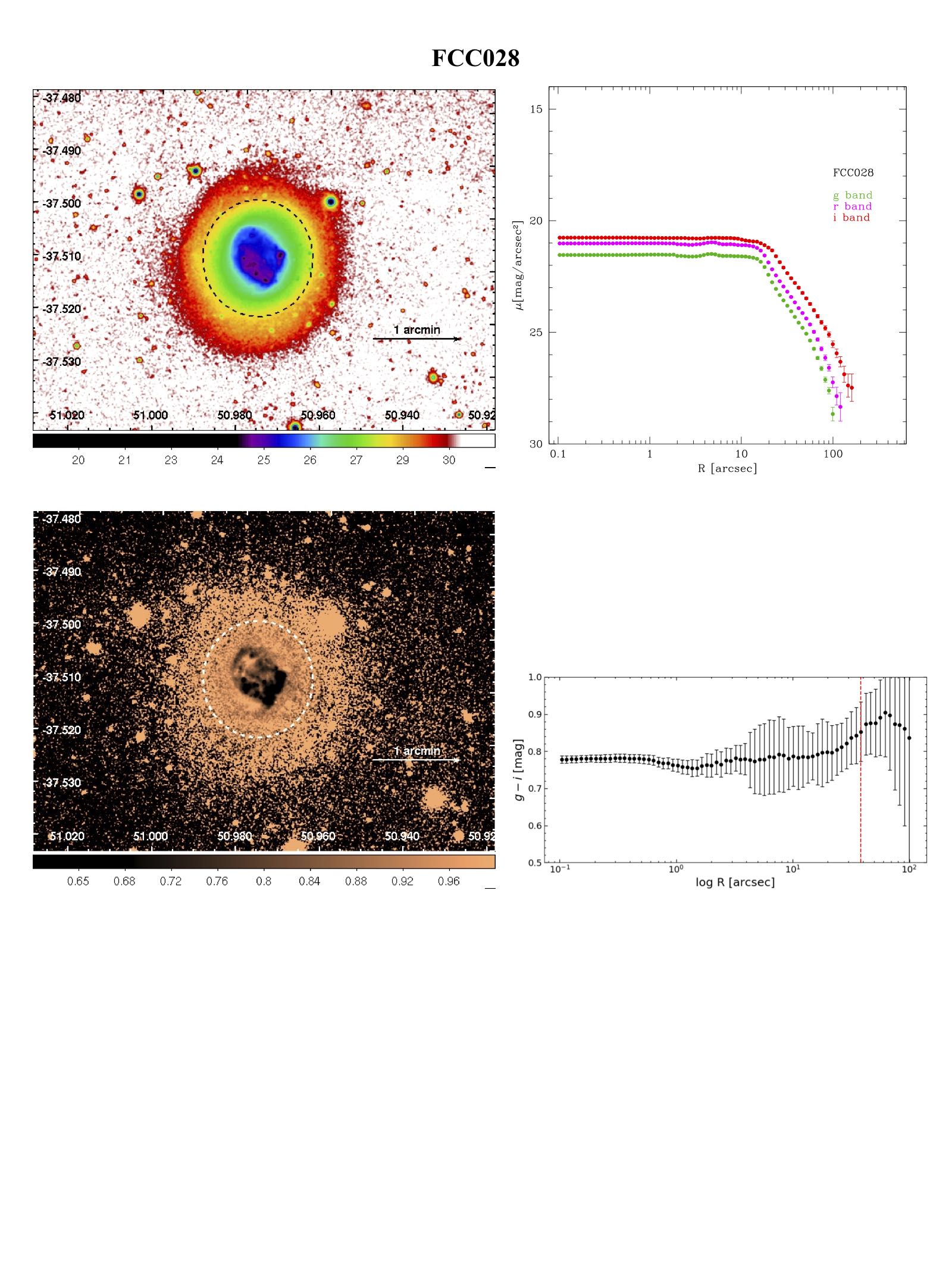}
     \vspace{-6.5cm}
     
      \caption{Surface Photometry of FCC028. Description same as Fig. \ref{FCC013}.}
             
         \label{FCC028}
   \end{figure*}

\begin{figure*}
  \centering
   \includegraphics[width=\hsize]{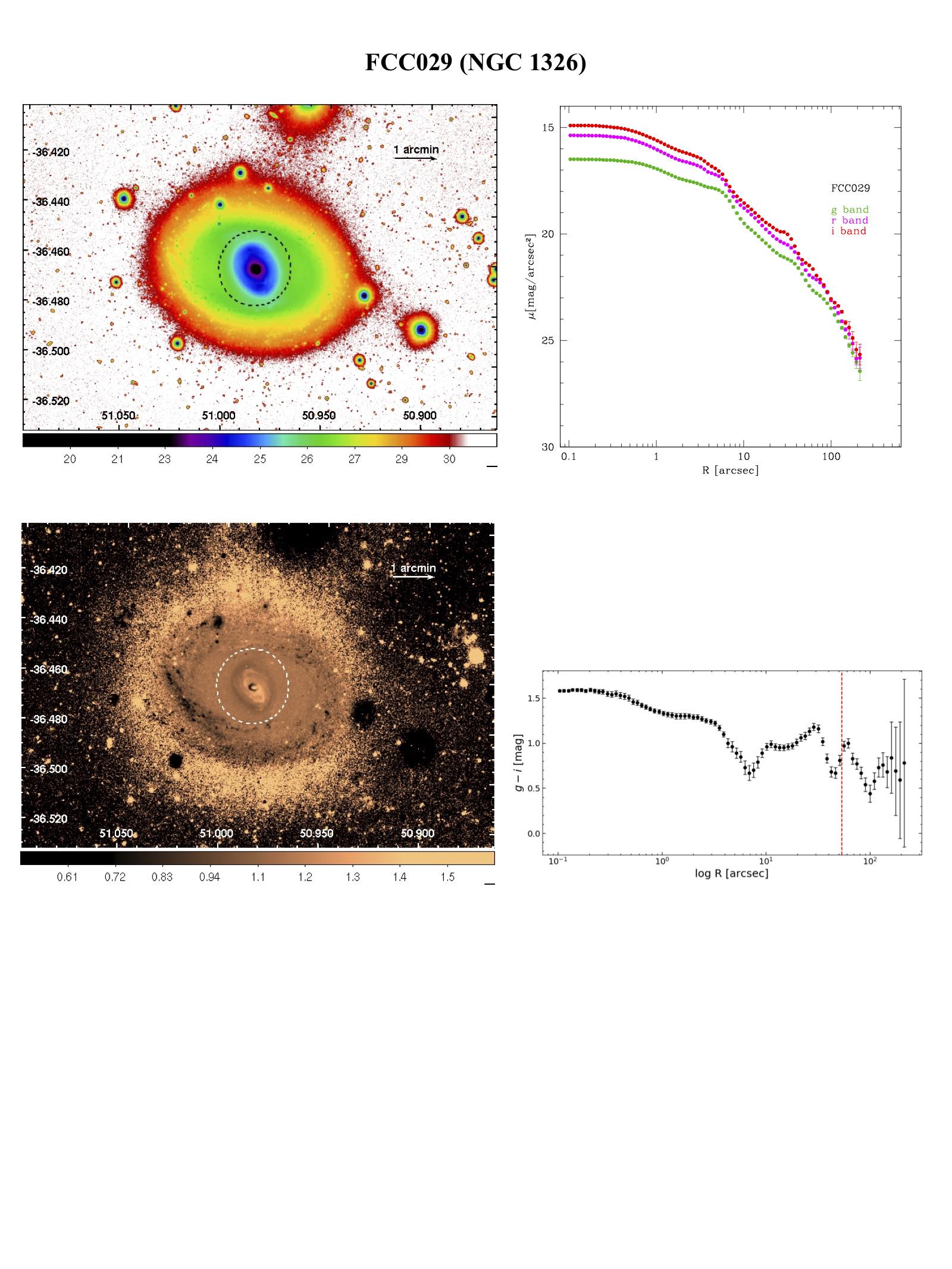}
     \vspace{-6.5cm}
     
      \caption{Surface Photometry of FCC029. Description same as Fig. \ref{FCC013}.}
             
         \label{FCC029}
   \end{figure*}

\begin{figure*}
  \centering
   \includegraphics[width=\hsize]{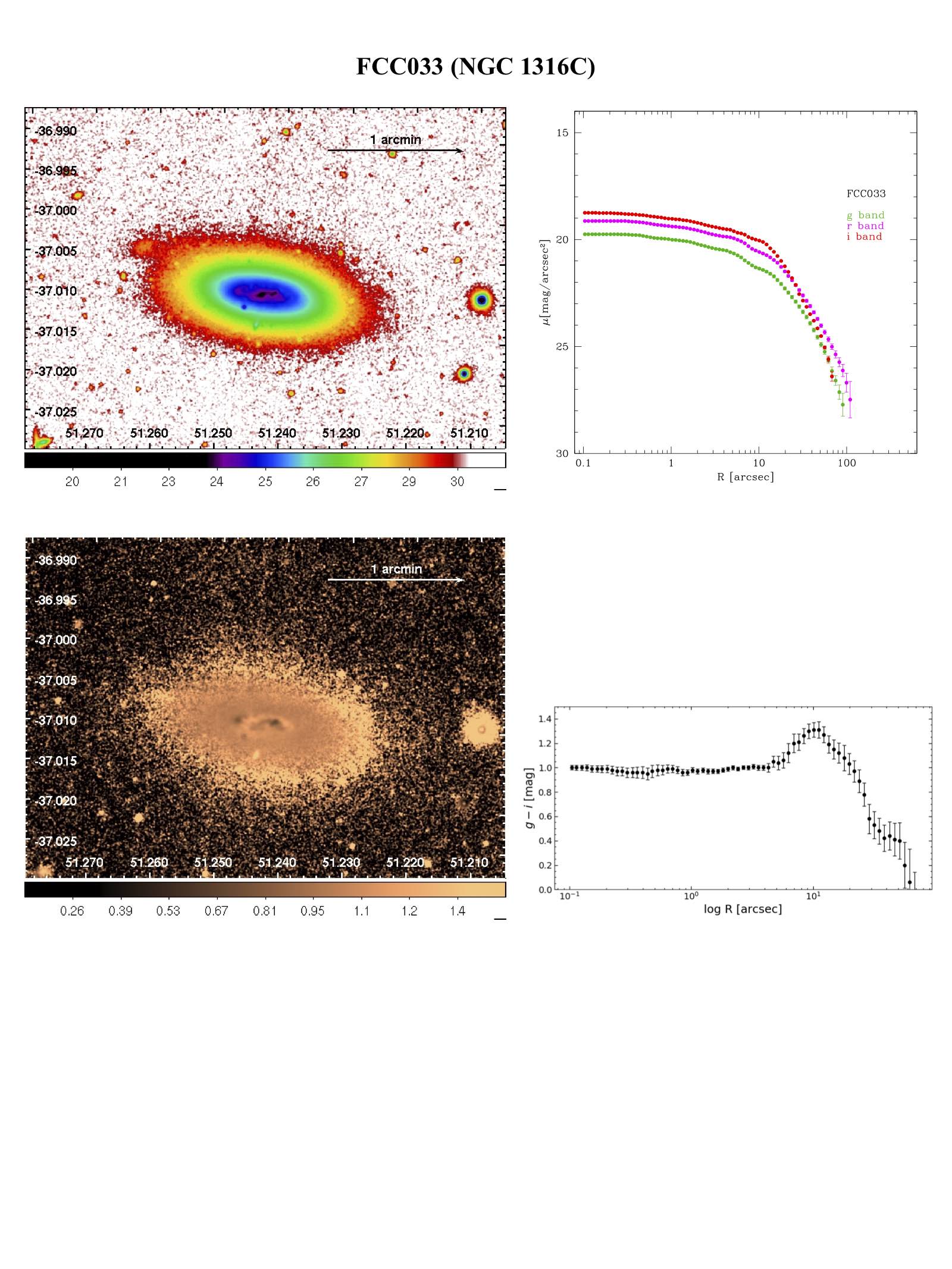}
       \vspace{-6.5cm}

      \caption{Surface Photometry of FCC033. Description same as Fig. \ref{FCC022}}.
             
         \label{FCC033}
   \end{figure*}
   
 \begin{figure*}
   \centering
   \includegraphics[width=\hsize]{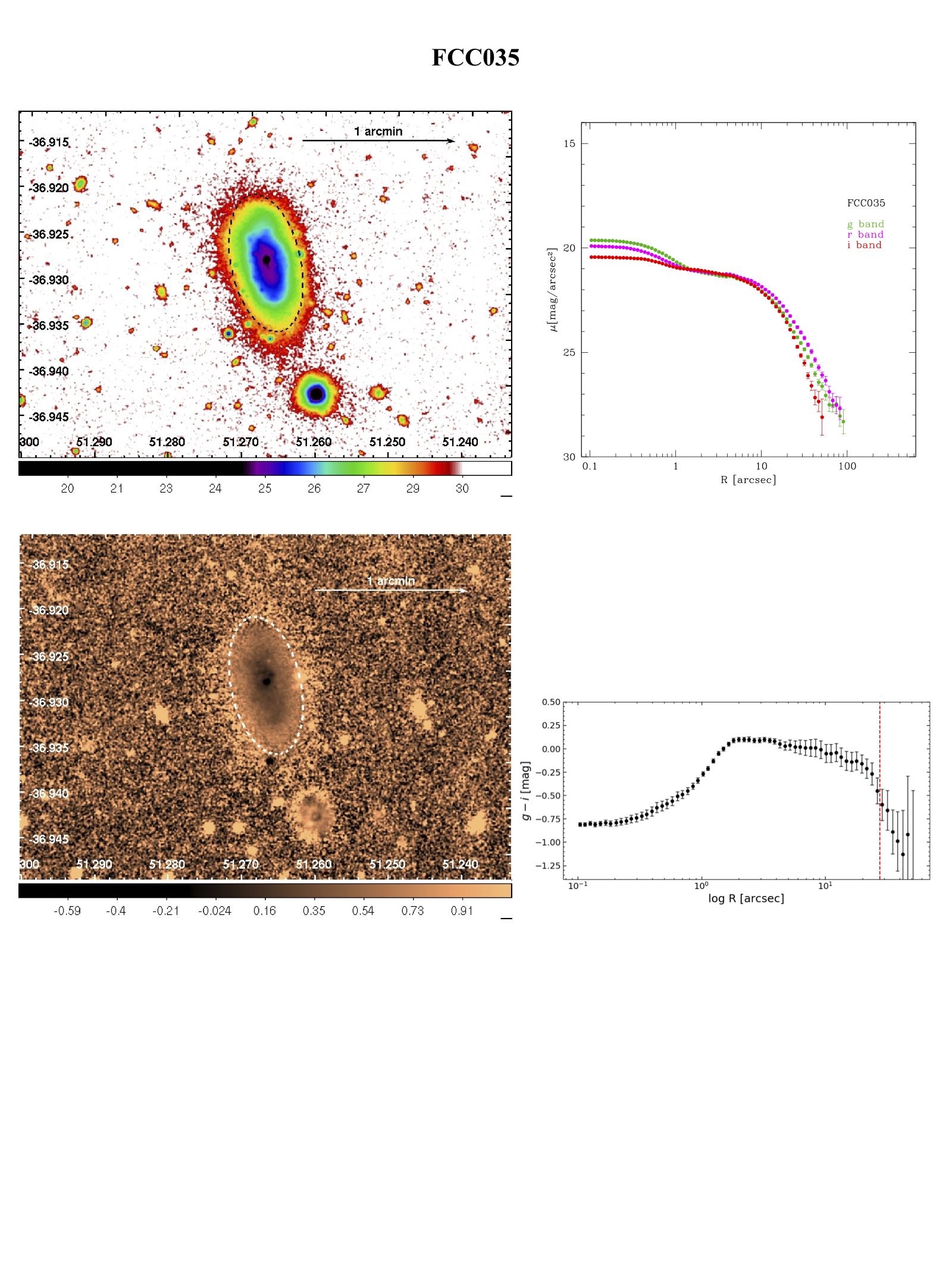}
    \vspace{-6.5cm}
    
      \caption{Surface Photometry of FCC035. Description same as Fig. \ref{FCC013}.}
       \label{FCC035}
      \end{figure*}

 \begin{figure*}
   \centering
   \includegraphics[width=\hsize]{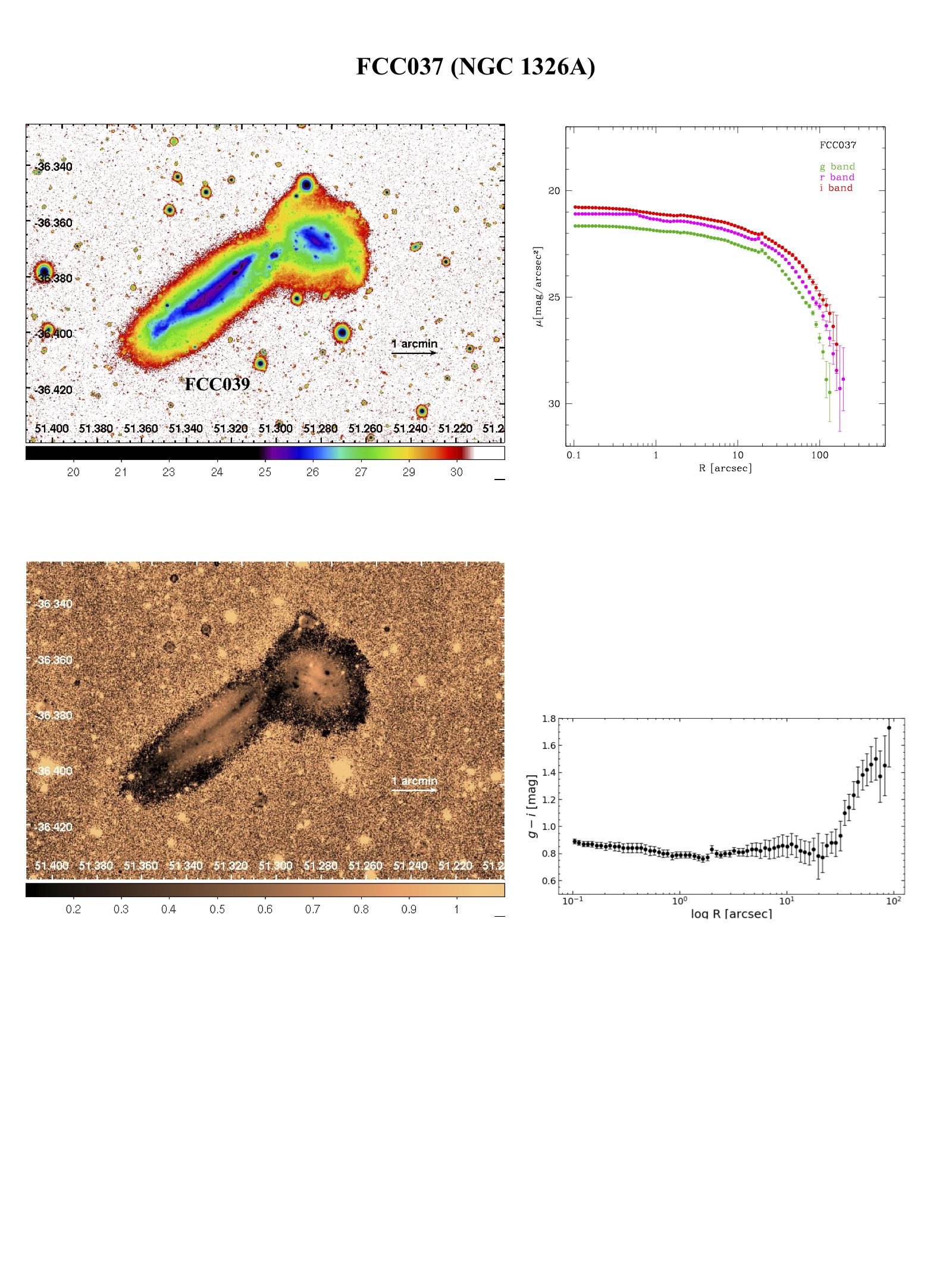}
   \vspace{-6.5cm}

      \caption{Surface Photometry of FCC037. Description same as Fig. \ref{FCC022}}.
     \label{FCC037}
   \end{figure*}

\begin{figure*}
   \centering
   \includegraphics[width=\hsize]{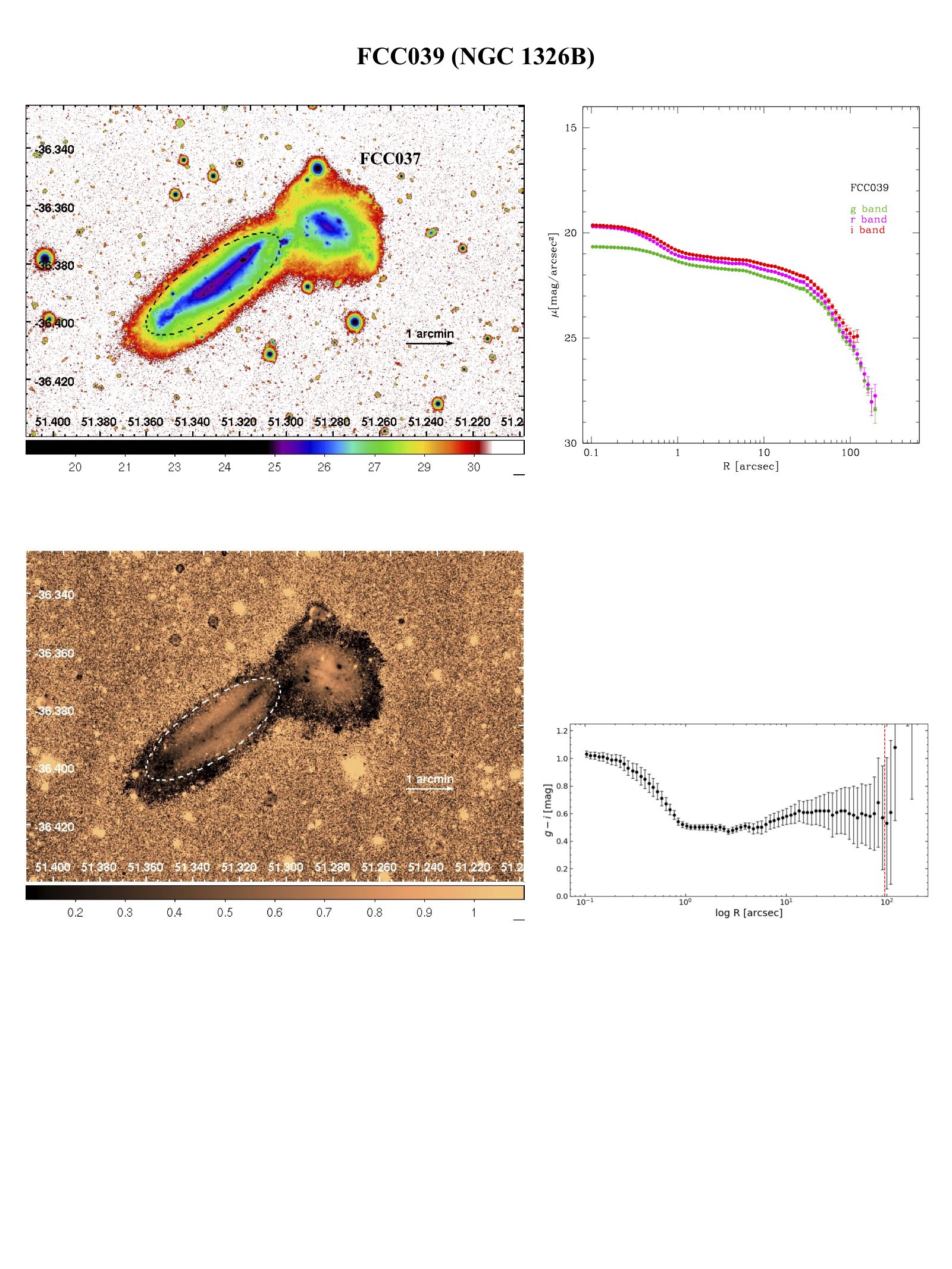}
    \vspace{-5.5cm}

      \caption{Surface Photometry of FCC039. Description same as Fig. \ref{FCC013}.}
             
         \label{FCC039}
   \end{figure*}

 \begin{figure*}
   \centering
   \includegraphics[width=\hsize]{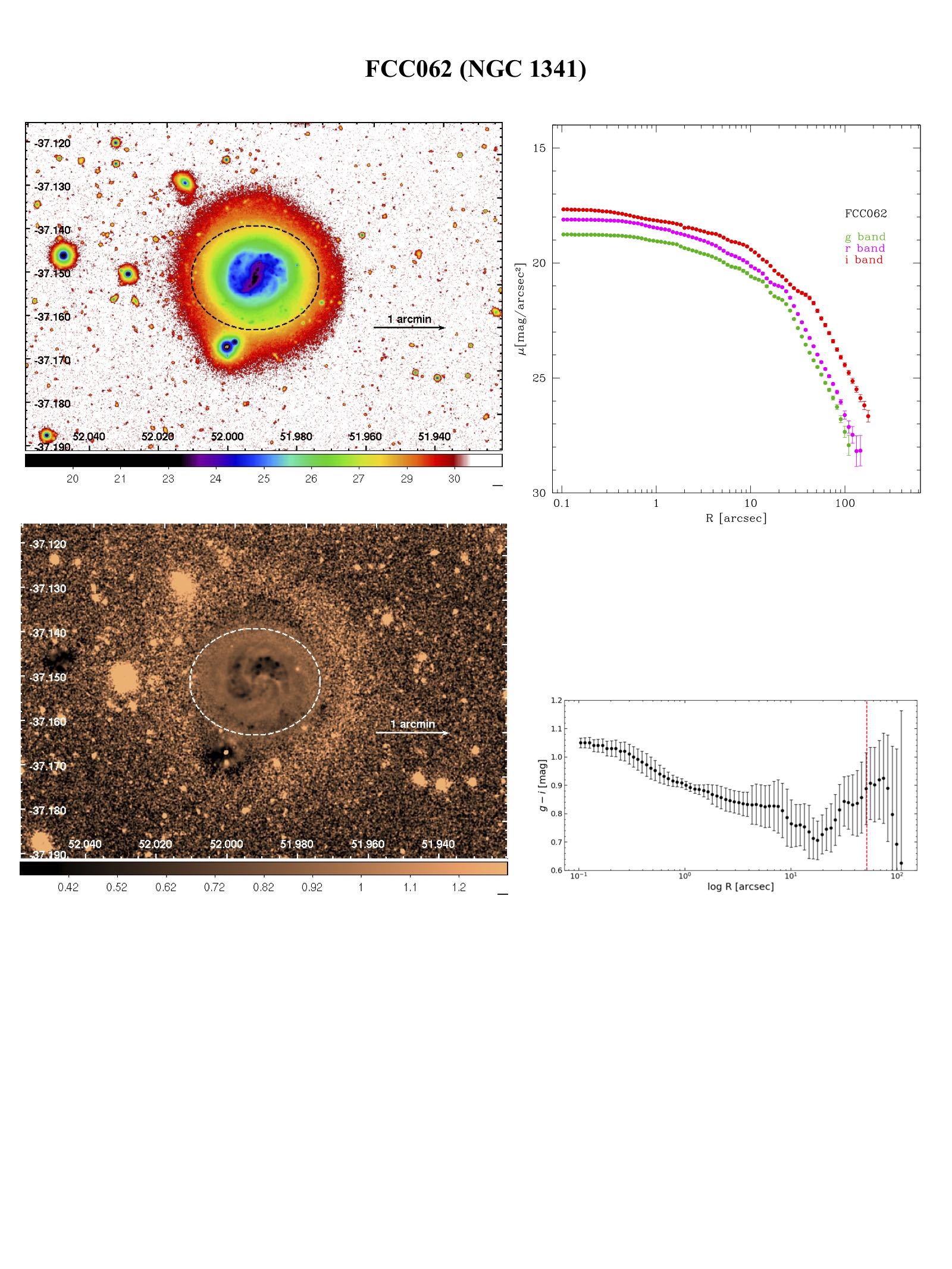}
    \vspace{-6.5cm}
    
      \caption{ Surface Photometry of FCC062. Description same as Fig. \ref{FCC013}.} 
      \label{FCC062}
   \end{figure*}
 \clearpage
 \onecolumn
 \begin{multicols}{2}
\section{Methodologies} \label{methods3}
\begin{figure*}[!h]
   \centering
   \includegraphics[width=\hsize]{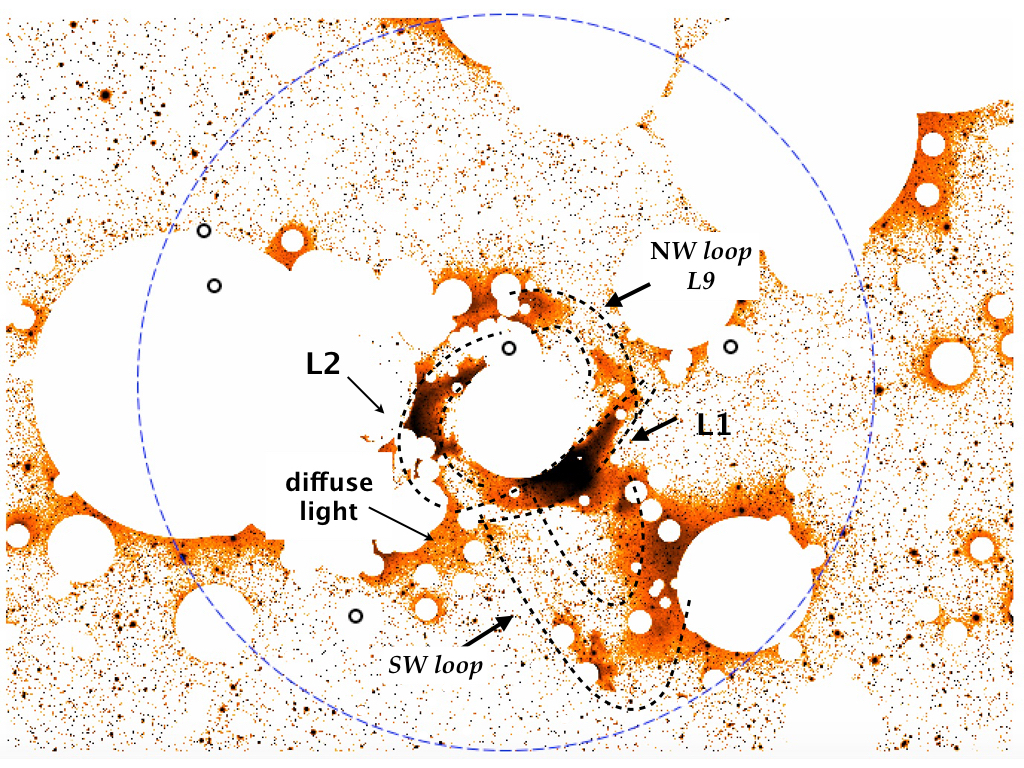}
    
      \caption{Residual image in $g$-band used for estimating the IGL. The image size is $\sim 87.7\times 63.4$ arcmin${^2}$. The blue dotted circle is the circular aperture with radius $=33.5^{\prime}$; the IGL is estimated in this region. All bright galaxies are marked with black circles. The known luminous structures- loops L1, L2, L5 (SW loop), L9 (NW loop)  are indicated (see also \citealt{Iodice2017}). A patch of diffuse light is also indicated.} 
      \label{igl_image}
   \end{figure*}

 In this section, we show the final residual image used in estimating the IGL in Fig.~\ref{igl_image}. We also briefly explain the multi-component fitting method we adopt, in Sect.~\ref{multi}. 
 The limiting radius of the intensity profiles in the $r$-band and its corresponding surface brightness are listed in Table~\ref{tab_lim}.
\end{multicols}

\begin{multicols}{2}
\subsection{Multi-component fits of the light distribution} \label{multi}
In this section, we briefly describe the 1D multi-component fitting method we adopt, published by \citet{spavone17,spavone20}. This is done to determine the starting point of the disc regions for the break-radius algorithm (described in Sect.~\ref{algor}).

\begin{figure*}[!h]
   \centering
   \includegraphics[width=17cm]{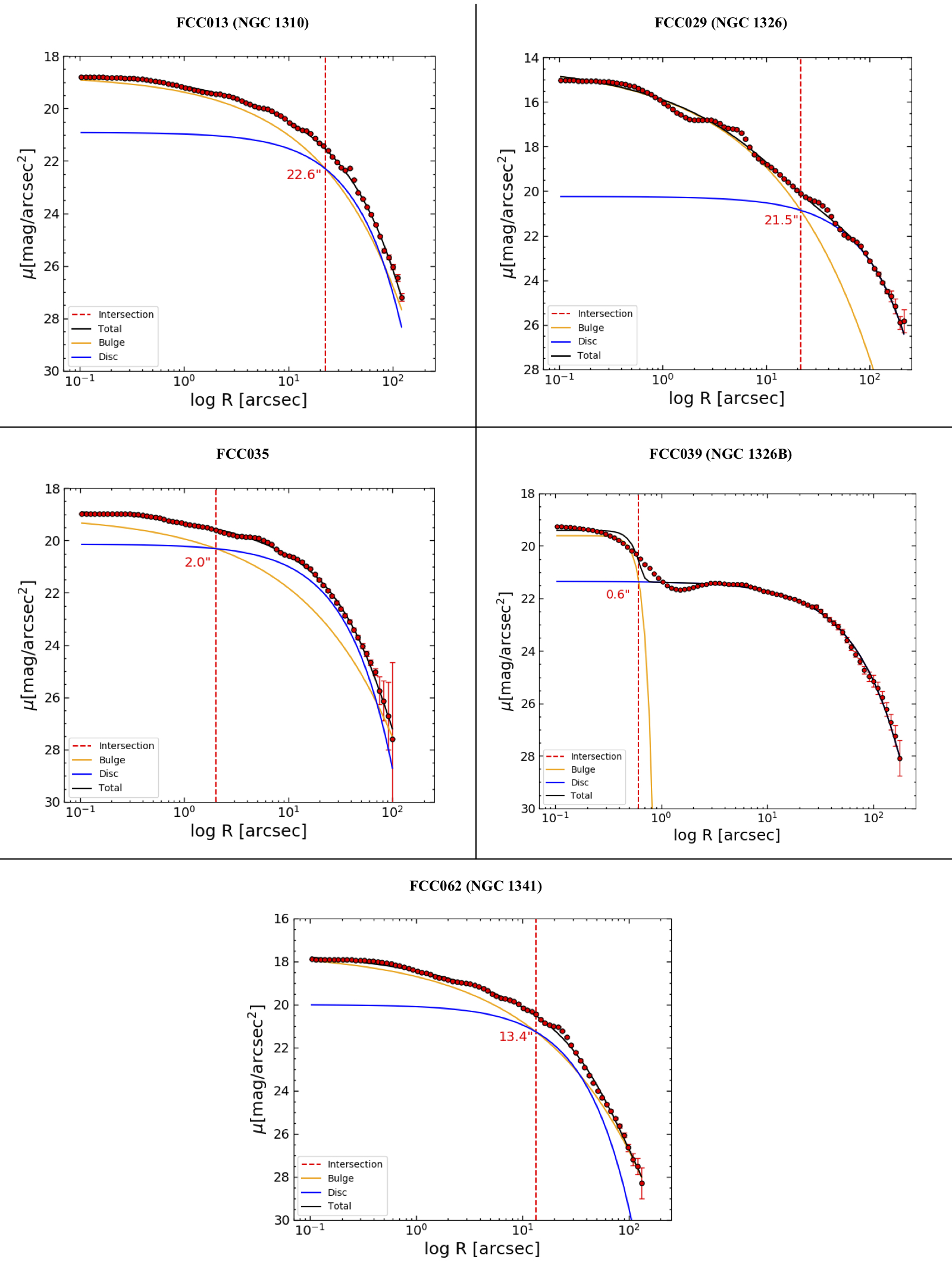}
      \caption{Multi-component fits of the surface brightness profiles ($r$-band) of galaxies with Type III disc break. In each plot, the multi-component fits, that is, bulge is indicated in orange, exponential disc is indicated in blue, and total (bulge+disc) is indicated in black. The intersecting point of the two components is indicated with a red dashed line.}
             \label{fits}
   \end{figure*}

We perform a multi-component fit to the deconvolved surface brightness profiles of five galaxies (NGC~1310, NGC~1326, FCC035, NGC~1326B, NGC~1341). 
The bulge component is fitted with a S{\'e}rsic law \citep{sersic63, caon93},
\begin{equation}
   \mu(R)= \mu_e + k(n)\left[\left(\frac{R}{r_e}\right)^{1/n}-1\right]
\end{equation}
where $k(n)= 2.17n+ 0.35$, $R$ is the galactocentric radius, $r_e, \mu_e$ are the effective radius and surface brightness. The disc component is fitted with an exponential profile ($n=1$),
\begin{equation}
    \mu(R)= \mu_0 + 1.086 \times R/r_h
\end{equation}
where $\mu_0$ and $r_h $ are the central surface brightness and exponential scale length.  The parameters are not fixed for the S{\'e}rsic component. The fitting procedure adopted here performs least-square fits using a Levenberg-Marquadt algorithm in which the function to be minimised is the rms scatter \citep[see][]{Seigar07}, defined as $\Delta= \frac{\sum_{i=1}^{n} \delta_i^2}{m} $  where $m$ is the number of data points and $\delta_i$ is $i^{\emph{th}}$ residual. The best fits are shown in Fig.~\ref{fits} and  their corresponding fitting parameters are listed in Table~\ref{best_fit}.

\end{multicols}
\onecolumn 
   
   \begin{table*}
\caption{Best-fitting parameters for five galaxies with detected disc-breaks in the Fornax~A subgroup.}              % title of Table
\label{best_fit}      % is used to refer this table in the text
\centering                                      % used for centering table
\begin{tabular}{lcccccc}          % centered columns (4 columns)
\hline\hline                        % inserts double horizontal lines
Object
& \multicolumn{3}{c}{Bulge}
& \multicolumn{2}{c}{Disc} 
& $r_{intersection}$
\\[+0.1cm]
 &\multicolumn{1}{c}{$\mu_e$}
 &\multicolumn{1}{c}{$r_e$} 
 &\multicolumn{1}{c}{$n$}                             
 
 & \multicolumn{1}{c}{$\mu_0$}
 & \multicolumn{1}{c}{$r_h$}
 &
\\[+0.1cm]
 
  & \multicolumn{1}{c}{[mag arcsec$^{-2}$]}
  &\multicolumn{1}{c}{[arcsec]}
  &\multicolumn{1}{c}{}
  &\multicolumn{1}{c}{[mag arcsec$^{-2}$]}
  &\multicolumn{1}{c}{[arcsec]}
  &[arcsec]\\
 (1) &\multicolumn{3}{c}{(2)}
  &\multicolumn{2}{c}{(3)}
   &\multicolumn{1}{c}{(4)} \\
 \hline\hline
FCC013 &22.34 $\pm$ 0.5 & 23.05 $\pm$ 1 & 1.83 $\pm$ 0.4& 20.91 $\pm$ 0.2 & 17.65 $\pm$ 3.1 & 22.6 \\
FCC029& 18.70 $\pm$ 0.2 & 8.96 $\pm$ 1 & 2.2 $\pm$ 0.1 & 20.24  $\pm$ 0.6 & 37.57$\pm$ 0.9&21.5 \\
FCC035 & 23.16 $\pm$ 0.9 & 22.47 $\pm$0.2 & 0.7 $\pm$ 0.5& 20.14 $\pm$ 0.7& 12.63 $\pm$ 0.1& 2.0 \\
FCC039 & 20.93 $\pm$ 0.3 & 0.49 $\pm$ 0.1& 0.65 $\pm$ 0.2& 21.26 $\pm$ 0.1& 27.1 $\pm$ 0.3& 0.6\\
FCC062 &21.9 $\pm$ 0.7& 19.92 $\pm$ 0.3& 2.21 $\pm$ 0.8& 20.0 $\pm$ 0.1& 11.56 $\pm$ 0.7& 13.4 \\
\hline
\end{tabular}
\tablefoot{Column 1 -- Fornax~A group members with detected disc-breaks; Column 2 -- Effective surface brightness, effective radius, and S{\'e}rsic index for the bulge component of each fit ; Column 3 -- Central surface brightness and scale length for the exponential component; Column 4 -- Radius at the intersection between the bulge and the disc component of the fits.}
\end{table*}
\begin{table*}
\caption{Limiting radius of intensity profiles.}              % title of Table
\label{tab_lim}      % is used to refer this table in the text
\centering                                      % used for centering table
\begin{tabular}{lcc}          % centered columns (4 columns)
\hline\hline                        % inserts double horizontal lines
Object & $R_{lim}$ & $\mu$ \\   
 & arcsec& mag arcsec$^{-2}$\\
  (1)&(2)&(3)\\
 \hline\hline
FCC013 (NGC~1310) &120.5 &27.07$\pm$ 0.3 \\
FCC022 (NGC~1317)&132.5 & 25.26$\pm$0.5\\
FCC028 &99.57& 27.24$\pm$0.2\\
FCC029 (NGC~1326)& 213.4& 25.82$\pm$0.5\\
FCC033 (NGC~1316C)&109.5  &  27.48$\pm$0.8\\
FCC035 & 82.29 & 27.68$\pm$0.5\\ 
FCC037 (NGC~1326A)&  145.8  &   27.66$\pm$0.5\\
FCC039 (NGC~1326B) &  160.4  &    27.22$\pm$0.4\\
FCC062 (NGC~1342)& 145.8    &   28.16$\pm$0.6\\
\hline
\end{tabular}
\tablefoot{Column 1 -- Fornax~A group members from \citet{Ferguson1989}; Column 2 -- Limiting radius in $r$-band ; Column 3 -- Surface brightness at the limiting radius.}
\end{table*}
   \end{appendix}
   
   \end{document}